# Ionization detail parameters and cluster dose: A mathematical model for selection of nanodosimetric quantities for use in treatment planning in charged particle radiotherapy

Bruce Faddegon, Eleanor A. Blakely, Lucas Burigo, Yair Censor, Ivana Dokic, Naoki Domínguez Kondo, Ramon Ortiz, José Ramos Méndez, Antoni Rucinski, Keith Schubert, Niklas Wahl and Reinhard Schulte

Corresponding author: Bruce Faddegon, University of California San Francisco, Department of Radiation Oncology 1600 Divisadero Street, San Francisco, CA 94143 USA, ORCID 0000-0002-4573-1582, bruce.faddegon@ucsf.edu, 415-548-0661

**Abstract.** Objective: To propose a mathematical model for applying Ionization Detail (ID), the detailed spatial distribution of ionization along a particle track, to proton and ion beam radiotherapy treatment planning (RTP).

Approach: Our model provides for selection of preferred ID parameters ($I_p$) for RTP, that associate closest to biological effects. Cluster dose is proposed to bridge the large gap between nanoscopic $I_p$ and macroscopic RTP. Selection of $I_p$ is demonstrated using published cell survival measurements for protons through argon, comparing results for nineteen $I_p$: $N_k, k = 2,3,...,10$, the number of ionizations in clusters of $k$ or more per particle, and $F_k, k = 1,2,...,10$, the number of clusters of $k$ or more per particle. We then describe application of the model to ID-based RTP and propose a path to clinical translation.

Main results: The preferred $I_p$ were $N_4$ and $F_5$ for aerobic cells, $N_5$ and $F_7$ for hypoxic cells. Significant differences were found in cell survival for beams having the same LET or the preferred $N_k$. Conversely, there was no significant difference for $F_5$ for aerobic cells and $F_7$ for hypoxic cells, regardless of ion beam atomic number or energy. Further, cells irradiated with the same cluster dose for these $I_p$ had the same cell survival. Based on these preliminary results and other compelling results in nanodosimetry, it is reasonable to assert that $I_p$ exist that are more closely associated with biological effects than current LET-based approaches and microdosimetric RBE-based models used in particle RTP. However, more biological variables such as cell line and cycle phase, as well as ion beam pulse structure and rate still need investigation.

Significance: Our model provides a practical means to select preferred $I_p$ from radiobiological data, and to convert $I_p$ to the macroscopic cluster dose for particle RTP.

Keywords: Nanodosimetry, ionization detail, track structure simulation, particle therapy, treatment planning, RBE

## 1. Introduction

Charged particle nanodosimetry and Monte Carlo track structure (MCTS) simulations are closely related methods that can link the number and density of energy transfers (ionizations and excitations) to the production of radiation-induced chemical species and biologically relevant DNA damage on the nanometer scale, as shown schematically in Figure 1 (Rucinski *et al* 2021). Nanodosimetry has focused on the number of ionizations created within a nanometer-scale volume representing a short DNA segment (1-2 helical turns or 10-20 base pairs) and the surrounding water layer. Due to the high scavenging capacity of the cellular environment, water radicals have a short lifetime, which means that only radicals created within a few nanometers from a DNA molecule have a reasonable chance to interact and create a DNA lesion (Michalik 1993). With increasing scavenging capacity, a larger fraction of DNA lesions is created by direct DNA ionization than to diffusing radicals. Ionization clusters created within the neighborhood of DNA segments will be responsible for DNA lesion clusters.

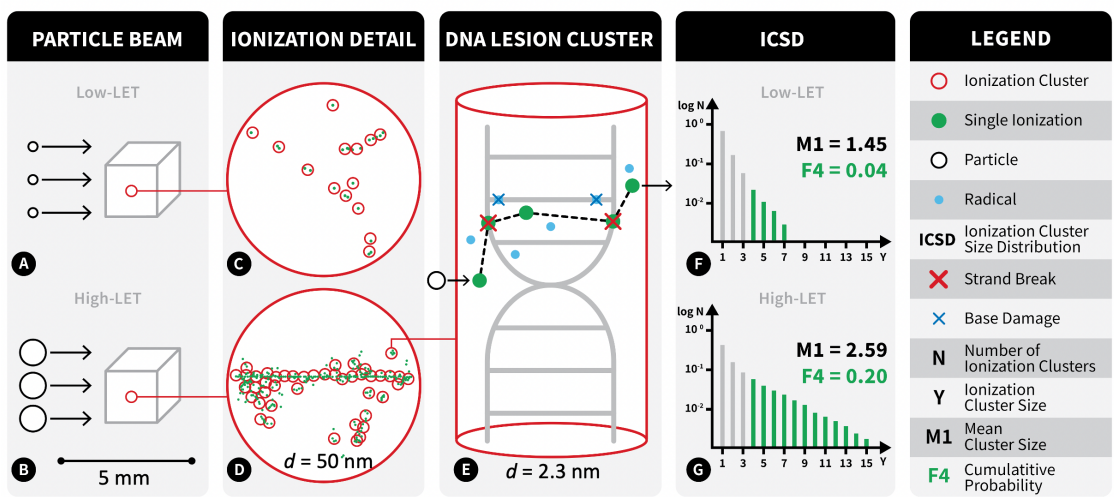

Figure 1: Schematic illustration of the relationship between a macroscopic particle radiation field and nanodosimetric quantities. Low-LET (Panel A) and high-LET (Panel B) radiation fields are incident on 2 mm width cubic regions, resulting in different spacing of ionization events, shown in 50 nm diameter spherical regions in Panels C and D, respectively. The high-LET particle leads to multiple ionizations in the 2.3 nm diameter cylinder shown in Panel E with the resulting ionization clusters leading to strand breaks (red x) and base damages (blue x) via free water radicals (blue dots) and direct DNA ionizations (green dots). Knowledge of the ionization detail allows quantification of ionization cluster size distributions (Panels F and G). Taken with permission, with revision, from (Rucinski *et al* 2021).

Biological effects predicted from nanodosimetric quantities present an opportunity for a new and more accurate paradigm for radiotherapy treatment planning (RTP) of particle radiotherapy with protons and heavier ions. Current proton RTP utilizes a constant relative biological effectiveness (RBE) value of 1.1, despite theoretical and experimental evidence demonstrating that the RBE increases across the spread-out Bragg peak (SOBP), with the steepest increase occurring towards the distal edge of the SOBP (Paganetti 2014). Clinical late effects at the distal edge of proton beams have recently been reported for several tumor sites, indicating the need to use a proton RBE value that is greater than 1.1 to accommodate uncertainties in RBE and in range (Scholz *et al* 1997, Inaniwa *et al* 2010, Zhang *et al* 2021, Deng *et al* 2021, Yang *et al* 2021, Lühr *et al* 2018).

Current carbon ion RTP is based on RBE models (Karger & Peschke 2018). The carbon doses that are isoeffective to photon doses are usually different for different RBE models, and currently no consensus exists among carbon ion treatment centers on which model to use (Fossati *et al* 2018). Recently, Nystrom *et al* (2020) stated: "We believe that the endless discussions of which is the most appropriate RBE-model and the exact values of the RBE for different tissues and in different parts of the dose distribution should be put on hold. Rather, efforts should be put in the development of clinical useful tools to visualize Linear Energy Transfer (LET) distributions and the possibility to include LET in the optimization of proton treatment plans".

New insights into the radiobiology of protons and ions demand a more sophisticated approach than current LET- and RBE-based RTP methods (McNamara *et al* 2019, Stasica *et al* 2020). These insights point toward the importance of ionization clusters causing complex DNA damage (DNA lesion clusters) (Falk & Hausmann 2020, Chatzipapas *et al* 2020, Hagiwara *et al* 2019, Nikitaki *et al* 2020, Wozny & Rodriguez-Lafrasse 2023). The most significant DNA lesion is a cluster of several double strand breaks (DSBs) (Charlton *et al* 1989). A complex double strand break (DSB) is repaired without error only by the homologous recombination repair pathway. This pathway is only available to the cell during late S and G2 cell cycle phases, and probably saturates after a few Gy of high-LET ion irradiation (Roobol *et al* 2020, Vignard *et al* 2013).

Ionization detail (ID) has been previously defined as the spatial distribution of ionization along a particle track (Ramos-Méndez *et al* 2018). The use of ID for proton and ion therapy planning is based on the premise that there exist certain ID parameters ($I_p$) that lead to similar biological responses when planned to be uniform across the target volume. For example, nanodosimetric event size probabilities per unit fluence can be related to interaction cross-sections for certain biological effects, e.g., cell inactivation (Blakely 1992, Conte *et al* 2017, Conte *et al* 2018, Rabus *et al* 2020, Conte *et al* 2023)

MCTS simulations combined with simulations of the subsequent DNA radiation chemistry demonstrate the association of nanodosimetric quantities with the production of complex DNA lesions (Schuemann *et al* 2018, Schuemann *et al* 2019). Prior research provides strong evidence (see, e.g., (Pinto *et al* 2002)) that the differential effects between high-LET and low-LET radiation at the cellular and molecular levels can be explained by local DNA lesion complexity. In fact, it was predicted almost 30 years ago that the detail of the spatial distribution of ionizing events along a particle track traversing the DNA molecule is closely correlated with the complexity and repairability of DSB in irradiated cells (Goodhead 1994).

*Nanodosimetric quantities in treatment planning for charged particle therapy* 4

In view of the information presented above, we have three goals for this paper: (1) To propose a mathematical model of ID that provides both a formal means to establish a quantitative dependence of biological effect on simplified ID parameters we call $I_p$ and a practical means of ID-based particle RTP. (2) To demonstrate the application of this model to contrast the association of LET and different $I_p$ with measured cellular effects by proton and ion beams. (3) To show how the ID model may be applied to ID-based RTP and propose a path to clinical translation.

We present a mathematical model of ID. While the physical quantities defined in the model presented here may be employed in radiobiological models, our model is not a radiobiology model. It is primarily intended to motivate the use of ID-based prescriptions in RTP. In order to employ the model, we propose that it first be used in conjunction with established RTP methods. As information, experience, and knowledge about ID parameters accumulates, and $I_p$ are found to associate strongly with biological effect, we anticipate our model will become more independent and will lead to RTP based exclusively on ID.

The approach proposed here is suitable to any ion used in therapy and includes the option of planning for mixed beam radiotherapy with a combination of different ions (see, e.g., Kopp *et al* 2020, Ebner *et al* 2021). Future multi-center trials with an ion therapy arm should include the means to compare treatment plans used by each institution (Lazar *et al* 2018, Roach *et al* 2016) and we anticpate that our ID-based approach will meet this challenge.

## 2. Materials and methods

In this section we describe mathematically the enumeration of ionization clusters produced in nanometeric volumes along a particle track. We show how to translate nanodosimetric quantities calculated from this mathematical model to the macroscopic scale of RTP, where dose and other physical quantities are calculated in millimeter-sized voxels. In the presentation of the mathematical model we adhere to mathematical principles while directing the presentation to physicists and biologists.

We first provide an overview of the mathematical model of ID. We then show how to derive $I_p$ from the absolute ionization cluster size frequency distribution, or the "frequency ICSD". Next, we introduce the cluster dose as a practical bridge between the nanoscopic $I_p$ and macroscopic ID-based RTP. Lastly, for demonstration purposes, we present one method to calculate frequency ICSD and use it to derive examples of $I_p$ and the associated cluster dose in macroscopic volumes to compare their association with biological effects.

Table 1 is a list of symbols used in the mathematical model. All entities in the table are defined in the sequel.

*Nanodosimetric quantities in treatment planning for charged particle therapy* 5

| | |
|---:|:---|
| $c$ | Particle class (particle type and energy) |
| $\mathscr{C}$ | A set of particle classes |
| $\nu$ | Ionization cluster size |
| $f_c(\nu)$ | Frequency distribution of $\nu$ for particle class $c$ |
| $I_p$ | ID parameter calculated from $f_c(\nu)$ |
| $I_p^c$ | ID parameter for particle class $c$ |
| $G_p$ | General operator to convert $f_c(\nu)$ to $I_p$ |
| $g$ | Cluster dose |
| $g_j^{(I_p)}$ | Cluster dose for $I_p$ in region $j$ |
| $\phi_j$ | Fluence in region $j$ |
| $\phi_j^c$ | Fluence of particles of class $c$ in region $j$ |
| $\mathscr{F}_j(\nu)$ | Track-length weighted frequency distribution in voxel $j$ |

Table 1: Key symbols used in the mathematical model.

## 2.1. A mathematical model of ID

The nanoscopic distribution of ionizing events along a particles track is described in detail by the particles ID. The aspects of ID most closely associated with biological effects are encapsulated by a set of numerical characteristics of this distribution labeled "ID parameters" and generically denoted by the symbol $I_p$. Each $I_p$ is calculated from the ID at the nanometer scale, as described below. These nanoscopic quantities are defined per particle, and are thus independent of particle fluence. In essence, $I_p$ are collapsed representations of the detailed spatial distribution of ionizations along particle tracks. Although these are currently calculated in water, $I_p$ in principle may be calculated on a DNA target, or on any molecular/cellular target. Examples of $I_p$ given at the end of Section 2.2 are used here to demonstrate $I_p$ calculation and selection for their association with biological effects.

For the purposes of RTP, $I_p$ is ideally chosen such that DNA damage in the nanometer-sized volume, and subsequent biological effects, are nearly the same for particles of different types and energies having the same $I_p$. That is, similar biological effects would result in regions that have a similar value of $I_p$. Neither the relationship between dose, dose rate, pulse structure and $I_p$ nor radiobiological differences in cellular response in tissues and organs, including different cell types, cell phase and inter- and extra-cellular environment are specifically discussed for the model presented here.

Our model allows the identification of $I_p$ that are strongly associated with biological effect. Potential $I_p$ are numerous, covering a range of possible mathematical forms to concisely represent the key aspects of the detailed distribution of ionization events for ID-based RTP.

In general, $I_p$ are determined in macroscopic-sized volumes. The term bin is used here for volumes with different shapes and sizes, such as a cylinder with the beam directed along a microscopic-sized axis or a several millimeter wide cube. In RTP, the

*Nanodosimetric quantities in treatment planning for charged particle therapy*     6

geometry and physical composition of the patient volume are represented by an image, which is discretized into a set of contiguous, millimeter-sized bins, such as derived from a CT scan. The term voxel is used here for a bin used in RTP. A word of caution regarding indexing of region, bin or voxel volumes: We use a single index $j$ so that our equations that include that index apply both to calculations done in bins and calculations done in voxels. This should not be ambiguous because it will be clear in each case which type of volume is considered.

For any $I_p$, we introduce the macroscopic quantity that we call "the generalized ionization cluster size dose," or simply cluster dose, denoted $g^{(I_p)}$, with units of 1/mass. As defined below, it is a fluence-weighted sum of the $I_p$ of the charged particles that result in ionization in a macroscopic bin. The cluster dose is a generalized quantity, its value depending on the generalized quantity $I_p$. Preferred $I_p$ define cluster dose quantities that are closely related to biological effects. Together, $I_p$ and $g^{(I_p)}$ are used to bridge the large gap in scale from nanometer-sized volumes used in nanodosimetry to macroscopic voxels used in RTP as depicted in Figure 2.

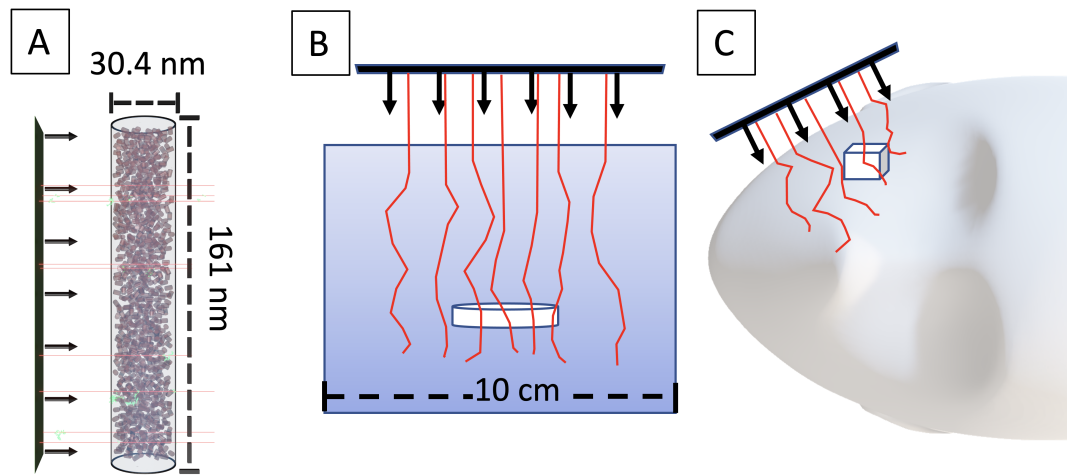

Figure 2: From nanoscopic to macroscopic: Regions, bins and voxels. The red lines depict particle tracks in the different geometries. A: Nanodosimetric quantities such as ID parameters are determined in nanometer-sized regions, as depicted in the cylinder shown on the left. B: Macroscopic quantities such as cluster dose are determined in millimeter-sized bins, such as the cylinder depicted in the middle. C: Millimeter-sized voxels, such as depicted in the patient on the right, are used in radiation treatment planning.

For the clinical dose range, inter-track effects on $g^{(I_p)}$ in proton and ion therapy at clinical doses and traditional (non-FLASH) dose rates are negligible. This is explained as follows. Dose in radiotherapy is low enough that the nanoscopic spatial distribution of ionized atoms in the initial physical stage of radiolysis is largely independent of dose or fluence. For example, for a dose of 2 Gy in treatment beams, in water, 5-20 MeV electron tracks are separated by ∼100 nm, 100 MeV proton tracks are separated by ∼240 nm, and low energy proton tracks in the Bragg peak are separated by ∼600 nm. Tracks of ions heavier than protons that deliver 2 Gy are separated even more. Therefore, $g^{(I_p)}$ is

*Nanodosimetric quantities in treatment planning for charged particle therapy* 7

to a very good approximation directly proportional to dose and fluence, given the same local spectra of particles in the bin where $g^{(I_p)}$ is determined, justifying the use of $I_p$ calculated from single-track ionization distributions.

### 2.2. The frequency distribution of ionization clusters and derived ID parameters

We define the integer parameter $\nu$ as a quantity that encapsulates the ID. This quantity, called "the ionization cluster size," is the number of ionizations in nanometer-sized regions in the vicinity of a particle track per unit track length. Probability distributions of $\nu$, or probabilistic ICSD, have been described before (Palmans *et al* 2015). Previous work on relating nanodosimetric quantities to biological effects (see, e.g., (Grosswendt *et al* 2007, Casiraghi & Schulte 2015, Bueno *et al* 2015, Ramos-Méndez *et al* 2018) has made use of quantities calculated from the ICSD.

In the model presented here, $I_p$ are calculated from absolute frequency distributions, not from probability distributions. We denote by class $c$ the physical characteristics of a single particle of a given type and energy. The value of any $I_p$ depends on the particle type and its energy and will be, thus, denoted in the sequel by $I_p^c$. We assume the existence of a frequency distribution function, for any given $c$, denoted by $f^c(\nu)$, i.e., $f^c : \mathbb{N} \to \mathbb{R}$ where $\mathbb{N}$ is the set of all integers and $\mathbb{R}$ is the real line (i.e., the $n$-dimensional Euclidean vector space $\mathbb{R}^n$ with $n = 1$).

Let $f^c(\nu)$ be the exact (non-stochastic) frequency distribution of $\nu$ for a single particle class $c$. MCTS calculations are commonly used to estimate $f^c(\nu)$. As an example, frequency ICSD are calculated using MCTS simulation provided in Section 2.5 using the source and geometry shown in Figure 2. In this case, a parallel beam of identical source particles uniformly irradiate a chromatin-fiber sized region. The average distance the particles travel in traversing the region is denoted $\ell$. The number of counts in each frequency bin is divided by the number of source particles and by $\ell$. The dimensions of $f^c(\nu)$ is length$^{-1}$.

Generally, a large number of source particles are simulated with MCTS for a specific source and scoring geometry to obtain $f^c(\nu)$ such that the calculation precision has minimal impact on the uncertainty of $I_p^c$ calculated from $f^c(\nu)$. However, this does not preclude the presence of systematic uncertainties, such as uncertainties in interaction cross-sections used in MCTS calculations (see, e.g., (Pietrzak *et al* 2021)). Frequency ICSD are known to vary with the choice of radiation source and sensitive volume geometry. For example, a larger target volume will increase the frequency of larger clusters.

We apply a general operator $G_p$ that converts frequency distributions, such as the frequency ICSD, to ID parameters. The mathematical model then applies to any method for extraction of $I_p$ from the frequency distribution. Examples are given shortly.

Let us denote by $\mathbb{F}$ the space of all frequency distributions $f^c$. Then $G_p : \mathbb{F} \to \mathbb{R}$ assigns to each frequency distribution a real value, which is the ID parameter $I_p^c$, and

*Nanodosimetric quantities in treatment planning for charged particle therapy*    8

we write
$$I_p^c := G_p[f^c]. \tag{1}$$

For RTP purposes, a set of $f^c(\nu)$ is pre-calculated with MCTS for a finite set of $c \in \mathscr{C}$ chosen to include all particle types that cover the range of energy of the charged particles that lay down tracks in the patient. The energy spacing should be sufficiently small for calculation of $I_p^c$ in a condensed history Monte Carlo (MC) simulation with negligible added error. The frequency distribution for a given particle energy may be interpolated or simply chosen as the one closest in energy to the energy of the particle.

This set of $f^c(\nu)$ may then be used to pre-calculate a set of $I_p^c$ values, $\{I_p^c \mid c \in \mathscr{C}\}$, calculated with Equation (1),

$$\{I_p^c \mid c \in \mathscr{C}\} = \{G_p[f^c] \mid c \in \mathscr{C}\}. \tag{2}$$

In the following we present two examples of $I_p$ definitions. Later we demonstrate how our mathematical model may be used to assess the degree of association of these example $I_p$ with cell survival data. The frequency ICSD $f^c(\nu)$ required for parameter calculation can be derived with MCTS simulations as described in Section 2.5. The summations are carried out over a range of cluster sizes.

The first $I_p$ example, $N_k^c$, defined for the particle class $c$, is the sum of cluster size frequencies weighted by cluster sizes $\nu \geq k$:

$$N_k^c := \sum_{\nu=k}^{\nu_{max}} \nu f^c(\nu). \tag{3}$$

Equation 3 gives the total number of ionizations in clusters of $k$ or more ionizations. This quantity is related to the average ionization cluster size.

The second $I_p$ example, $F_k^c$, defined for the particle class $c$, is the sum of cluster size frequencies with $\nu \geq k$:

$$F_k^c := \sum_{\nu=k}^{\nu_{max}} f^c(\nu). \tag{4}$$

These example definitions should not be confused with theoretical quantities derived from probabilistic ICSD (Palmans *et al* 2015). The operator definition of $I_p$ parameters could be more complicated than the examples shown above. For example, ID-based RTP of a mixed beam of protons and carbon ions might be better planned using an $I_p$ with a linear combination of two or more $F_k$ rather than using a single $F_k$ such as $F_5$ alone. The selection of preferred $I_p$, including combined $I_p$, is left for future research.

### 2.3. Nanodosimetric parameters calculated on the macroscale

Next we consider a practical means of ID-based RTP using $I_p$. The parameters we consider may be determined in bins or voxels. We use the index $j$ to identify the bin or voxel in the phantom or patient geometry.

*Nanodosimetric quantities in treatment planning for charged particle therapy* 9

Our approach is based on two fundamental requirements. Firstly, that the $I_p$ of the portions of the tracks (track segments) laid down by individual particles in a given bin $j$ that correspond to the same particle class $c$ may be summed together; that is, there is no consideration of overlap between particle tracks. This may be an important limitation at ultra-high dose rates. Secondly, that for each $c \in \mathscr{C}$, the sum of $I_p^c$ is linearly related to the cumulative length $t_j^c$ of all such track segments in bin $j$ (such as in voxel $j$ if a voxel is considered as a bin for RTP purposes). To be clear, the length of the track segment of a particle of a specific class in a given volume is the intersection of the volume with the portion of the track with the particle in the energy range corresponding to the frequency ICSD or the derived ID parameter for that particle class.

The density of ionization scales with the density of the medium. To account for this, the track length is scaled by density in $g_j^{(I_p)}$. Then, to equate this quantity in bins of different density, we scale by the reciprocal of the mass of the medium. Together, this results in scaling by the reciprocal of the volume of the bin.

We define the cluster dose $g_j^{(I_p)}$ for the ID parameter $I_p$ as the sum over the track segments in bin $j$ of volume $V_j$ as follows:

$$g_j^{(I_p)} := \frac{1}{m_j} \sum_{c \in \mathscr{C}_j} \frac{\rho_j}{\rho_0} t_j^c I_p^c = \frac{1}{\rho_0 V_j} \sum_{c \in \mathscr{C}_j} t_j^c I_p^c, \qquad (5)$$

where the mass of the material in the bin is the product of the bin volume and density, $m_j = V_j \rho_j$, $t_j^c$ is the cumulative track segment length of particles of class $c$ (type and energy) in bin $j$, $\rho_0$ is the density of the medium used for the calculation of $I_p^c$ for all $c \in \mathscr{C}_j$ (generally water of 1 g/cm$^3$), and $\mathscr{C}_j$ is the finite set of all charged particles that induced ionizations in bin $j$. The track length of the source particle of class $c$ in the MCTS calculation of the frequency distribution, $\ell$, is included in $I_p^c$.

By definition, $I_p$ is independent of fluence. Conversely, the portion of $g_j^{(I_p)}$ corresponding to particle class $c$, is linearly proportional to the summed length of track segments of particles of class $c$ in the volume.

In voxels of dimensions typically used in treatment planning, individual charged particles either traverse the voxel, are set in motion in the voxel (secondaries), and/or stop in the voxel. A particle may lose enough energy while laying down its track in the voxel to move into the energy range of a different particle class, with a different value of $I_p$ derived from a different frequency ICSD. To account for this in the definition of $g_j^{(I_p)}$, the track of each particle in voxel $j$ is subdivided into shorter track segments that coincide to a particular particle class. The lengths of the track segments laid down by particles of a particular class are then summed together.

The track lengths are scaled with the density of the material (tissue) in the volume compared to the density of the medium in the MCTS simulations used to calculate the frequency ICSD for each particle class. That is, density scaling is invoked to account for the linear increase in $g_j^{(I_p)}$ with density, requiring that the medium composition is the same. To account for medium composition, the frequency ICSD may be determined for each medium. However, MCTS simulation is currently limited in the medium

*Nanodosimetric quantities in treatment planning for charged particle therapy* 10

composition, generally to liquid water, due to limitation in the available cross-sections at this scale.

For $F_k$ from Equation (4) selected as the ID parameter, for example, we have

$$g_j^{(F_k)} = \frac{1}{\rho_0 V_j} \sum_{c \in \mathscr{C}_j} t_j^c F_k^c. \tag{6}$$

In practice, Equation (6) can be applied in condensed history MC simulation by summing up the product of pre-calculated $F_k$ with the length of the track segment of particles of each class in bin $j$.

Comparing proton beams to carbon ion beams, the values of $F_4$ for a 100 MeV proton and a 100 MeV/u carbon ion are 0.0053 and 0.31 respectively, calculated using the pre-calculated frequency ICSD. To yield the same density of clusters with four or more ionizations, $g_j^{(F_4)}$, it would take 58 times more protons than carbon ions. Even if an $I_p$ associates exactly with biological effects, the biological effects may be somewhat different for a proton beam with the same $g_j^{(I_p)}$ as a carbon ion beam. This is considered further in the Section 2.4. Which $I_p$ best associate with biological effects remains to be determined.

Next we focus our attention on the operator $G_p$. We assume that the space $\mathbb{F}$ of all frequency distributions $f^c$ is a linear space, and that the operator $G_p$ is linear on $\mathbb{F}$. This means that for any finite set of frequency distributions $\{f^c\}_{c \in \mathscr{C}}$ belonging to $\mathbb{F}$ and any finite set $\{\alpha_c\}_{c \in \mathscr{C}}$ of real numbers, with $\alpha_c$ a weighting factor applied to the frequency ICSD, we have $\sum_{c \in \mathscr{C}} \alpha_c f^c \in \mathbb{F}$, and

$$G_p \left[ \sum_{c \in \mathscr{C}} \alpha_c f^c \right] = \sum_{c \in \mathscr{C}} \alpha_c G_p[f^c]. \tag{7}$$

Then, from Equations (1), (5) and (7), employing $\alpha_c = t_j^c$, for all $c \in \mathscr{C}$, we have

$$g_j^{(I_p)} = \frac{1}{\rho_0 V_j} \sum_{c \in \mathscr{C}_j} t_j^c G_p[f^c] = \frac{1}{\rho_0 V_j} G_p \left[ \sum_{c \in \mathscr{C}_j} t_j^c f^c \right]. \tag{8}$$

We introduce the dimensionless track-length weighted frequency distributions $\mathscr{F}_j(\nu)$ in bin $j$. These are dimensionless quantities. For any bin $j$ and for any $\nu \in \mathbb{N}$, $\mathscr{F}_j : \mathbb{F} \to \mathbb{R}$ is defined by

$$\mathscr{F}_j(\nu) := \sum_{c \in \mathscr{C}_j} t_j^c f^c(\nu). \tag{9}$$

Then, the linearity of $G_p$ in Equation (7) and the additivity assumption in Equation (9) yield

$$g_j^{(I_p)} = \frac{G_p[\mathscr{F}_j]}{\rho_0 V_j}. \tag{10}$$

Therefore, due to our linearity assumption on $G_p$, Equations (5) and (10) are equivalent and either of them may be used to calculate $g_j^{(I_p)}$.

*Nanodosimetric quantities in treatment planning for charged particle therapy* 11

For example, if $f(\nu)$ is the frequency ICSD and the operator $G_p$ maps $f(\nu)$ to $F_k$, we can use the following equation, instead of Equation (6):

$$g_j^{(F_k)} = \frac{1}{\rho_0 V_j} \sum_{\nu=k}^{\nu_{max}} \mathcal{F}_j(\nu). \tag{11}$$

In practice, Equation (11) can be applied in condensed history MC simulation by summing up the product of frequency ICSD pre-calculated with MCTS simulation with the length of the track segment of each particle in a voxel (Equation (9)). The result of Equation (6) and Equation (11) will be identical as long the same frequency ICSD are used.

### 2.4. Additional considerations for ID-based RTP

The use of ID parameters in RTP rests on the premise that there exist preferred $I_p$ such that charged particle beams with the same value of $g_j^{(I_p)}$ in different voxels have nearly the same biological response per unit fluence, regardless of the charged particle composition in the voxels. Which $I_p$ better associate with biological effects, independent of LET, is currently unknown and needs to be determined experimentally.

From Equation (5), for a set $\mathscr{C}_j$ of charged particle types and energies in voxel $j$, and a set $\mathscr{C}_j'$ consisting of a different mixture of charged particles in the same voxel, $\mathscr{C}_j$ and $\mathscr{C}_j'$ can yield the same biological effects when their $g_j^{(I_p)}$ are equal, that is, under the condition that

$$\sum_{c \in \mathscr{C}_j} t_j^c I_p^c = \sum_{c \in \mathscr{C}_j'} t_j^c I_p^c. \tag{12}$$

The same applies for two different voxels $j$ and $k$ under the condition that

$$\frac{1}{V_j} \sum_{c \in \mathscr{C}_j} t_j^c I_p^c = \frac{1}{V_k} \sum_{c \in \mathscr{C}_k} t_k^c I_p^c. \tag{13}$$

That is, given an $I_p$ such that particle classes having the same $I_p$ have the same biological effect, then under the condition of Equation (13), voxels with the same $g_j^{(I_p)}$ will have the same biological effect.

Equations (12) and (13) may be satisfied without requiring that the fluence be the same. For example, a high fluence of low $I_p$ particles (such as protons) passing through a voxel could yield the same track-length-weighted sum of $I_p$ as a much lower fluence of high $I_p$ particles (such as carbon ions). Although this results in the same $g_j^{(I_p)}$ values, as eluded to earlier for $F_4$ following Equation (6), the biological effects may not be the same when radiation fields of very different quality are used.

In the following, we present a more conservative approach. We first define a fluence weighted average of $I_p$ for a mixed field $\mathscr{C}_j$ of particles in voxel $j$:

$$I_p^{\mathscr{C}_j} := \frac{\sum_{c \in \mathscr{C}_j} \phi_j^c I_p^c}{\sum_{c \in \mathscr{C}_j} \phi_j^c}. \tag{14}$$

*Nanodosimetric quantities in treatment planning for charged particle therapy* 12

Since the fluence of particles of class $c$ in voxel $j$, denoted by $\phi_j^c$, is related to the path length $t_j^c$, we may replace $\phi_j^c$ by

$$\phi_j^c = \frac{t_j^c}{V_j}, \tag{15}$$

which leads to

$$I_p^{\mathscr{C}_j} = \frac{\sum_{c \in \mathscr{C}_j} t_j^c I_p^c}{\sum_{c \in \mathscr{C}_j} t_j^c}. \tag{16}$$

Replacing $I_p^c$ by $f^c$ in Equation (14), which can be achieved by using the identity operator for $G_p$ in Equation (1), we can define a fluence weighted average of $f(\nu)$ for a mixed field of particles in voxel $j$:

$$f^{\mathscr{C}_j}(\nu) := \frac{\sum_{c \in \mathscr{C}_j} \phi_j^c f^c(\nu)}{\sum_{c \in \mathscr{C}_j} \phi_j^c} = \frac{\sum_{c \in \mathscr{C}_j} t_j^c f^c(\nu)}{\sum_{c \in \mathscr{C}_j} t_j^c}. \tag{17}$$

As for the $I_p$ and frequency ICSD in Equation (1) their related fluence-averaged quantities in Equations (14) and (17) are independent of fluence.

Using the linear operator $G_p$, an alternative expression for $I_p^{\mathscr{C}_j}$ is:

$$I_p^{\mathscr{C}_j} = G_p[f^{\mathscr{C}_j}]. \tag{18}$$

This equation is useful for discriminating between different ID parameters on the basis of their association with biological effects (see Section 2.6) and for ID-based RTP optimization (see Section 4.5).

The total fluence of particles in voxel $j$ is:

$$\phi_j := \sum_{c \in \mathscr{C}_j} \phi_j^c. \tag{19}$$

Combining Equation (5) with Equations (15), (16) and (19) leads to:

$$g_j^{(I_p)} = \phi_j I_p^{\mathscr{C}_j} / \rho_o \tag{20}$$

It is apparent from this equation that a uniform $I_p^{\mathscr{C}_j}$ in the target volume will result with a uniform $g_j^{(I_p)}$ if the local fluence $\phi_j$ is also uniform. The requirement of locally uniform $I_p^{\mathscr{C}_j}$ and fluence $\phi_j$ is more restrictive than a locally uniform $g_j^{(I_p)}$ alone. The latter removes the requirement of a locally uniform fluence.

This completes our presentation of the mathematical model of ID.

## 2.5. Calculation of a database of frequency ICSD

In the remainder of the Materials and Methods section we present methods pertaining specifically to the examples we have chosen to demonstrate practical applications of the ID model. We first consider calculation of the frequency distributions $f^c(\nu)$. A set of these distributions were pre-calculated to cover the range of charged particle types and energies (represented by the finite set of particle classes $\mathscr{C}$) in irradiated regions.

*Nanodosimetric quantities in treatment planning for charged particle therapy*     13

We used TOPAS-nBio (Schuemann *et al* 2019) to perform MCTS simulations to calculate a database of frequency ICSD following published procedures (Ramos-Méndez *et al* 2018). The pre-calculated database included all charged particle types present in ion therapy from proton through carbon ions. The set contained frequency ICSD for electrons with energies from 10 keV-1 MeV, protons with energies ranging from 0.5–100 MeV, 4-He ions with energies from 1–100 MeV/u, 7-Li, 9-Be, 11-B, 12-C, 14-N, 16-O ions with energies from 1–1000 MeV/u, and 19-F, 20-Ne, 23-Na, 24-Mg, 27-Al, 28-Si, 32-S, 35-Cl and 40-Ar with energies ranging from 1-1000 MeV/u. Units of MeV were used for electrons and protons, MeV/u for ions heavier than protons. The lowest energy for each particle type in the database was set to the minimum energy required for the particle to traverse the 30.4 nm diameter encompassing cylinder or the minimum energy that could be simulated in Geant4-DNA, whichever was greater.

The set of frequency ICSD calculated for each particle consisted of 100 energy bins, evenly-spaced on the logarithmic scale. We used the latest models (option4 models) for charged particle transport recommended by the Geant4-DNA collaboration (Incerti *et al* 2018). The source and geometry used to calculate the frequency ICSD was chosen for demonstration purposes and should not be taken as a preference over other possibilities.

A 161 nm long, 30.4 nm diameter cylinder was placed at the center of a 100 nm × 200 nm × 100 nm rectangular parallelopiped "world" volume of water. This cylinder, approximating a chromatin fiber, was oriented with its axis along the 200 nm length of the world volume. A 161 nm long × 30.4 nm source was normally incident on one of the 200 nm length surfaces of the world volume, positioned to match the cross-section of enclosed cylinder.

The cylinder encompassed a large number of cylindrical target volumes, each of 2.3 nm diameter and 3.4 nm length, representing short DNA segments of 10 base pairs (Figure 2A). The target volumes were randomly oriented and randomly placed without overlap throughout the encompassing cylinder (Ramos-Méndez *et al* 2018). The cylinders were placed using the G4RandFlat function. If a cylinder overlapped another, it was deleted and resampled. A total of 1800 target volumes were placed in the encompassing cylinder, close to the maximum that could be achieved. In sampling of ion tracks of each particle class, the average track length through the encompassing cylinder, $\ell$, was assumed to be the mean chord length of the cross-sectional area of the cylinder of 23.88 nm. This ignores any scattering of the primary particle. These simulations provided us with look-up tables of $f^c(\nu)$ for the entire set $\mathscr{C}$ of particle classes.

2.6. *Association of ID parameters with biological effect*

As a demonstration of the practical use of our model, we compared $g_j^{(I_p)}$ to cell survival. In radiobiology experiments and in radiotherapy with radiotherapy quality beams, the characteristics of the source (energy, spectral width, etc.) result in a radiation field in the macroscopic volume element of interest (bin or voxel) consisting of a mix of primary and secondary particles of different types and energies. In this case, the fluence (or track segment length) of particles of each class is needed in order to calculate $g_j^{(I_p)}$. The fluence may be calculated with condensed history MC simulation. We demonstrate examples of $I_p$ association with published cell survival data from two separate sets of experiments.

In the first set (Blakely *et al* 1979), survival of human kidney T-1 cells grown in monolayers on thin glass coverslips was measured at various points and residual energies along the pristine, passively scattered Bragg curve of carbon, neon and argon beams with initial energies of 400, 425 and 570 MeV/u, respectively. Measurements were made at LBNL using the BEVALAC accelerator (Curtis 1974). In the second set (Dokic *et al* 2016), survival of human alveolar adenocarcinoma A549 cells grown in monolayers was measured at the center of a scanned 1 cm spread out Bragg peak (SOBP) of clinical proton, helium, carbon and oxygen beams. Measurements were made at the Heidelberg Ion Therapy (HIT) facility (Combs *et al* 2010, Jäkel *et al* 2022).

Simulations of these beams were done with the condensed history MC method using TOPAS (Perl *et al* 2012, Faddegon *et al* 2020). The geometry used to simulate the setup of the experiments performed at LBNL (Blakely *et al* 1979) consisted of a water cylinder of 30 cm length and 20 cm diameter. Ion beams of 4 cm radius were normally incident on the central axis of one side of the cylinder.

The cylinder was divided into 3000 depth bins of 0.01 cm each. Fluence, dose, and frequency ICSD were scored in 10 cm radius cylindrical bins in each depth bin. The scoring bins were thus 0.01 cm thick (in depth), 10 cm radius cylindrical bins. This calculation takes advantage of the Geometry Equivalence Theorem (Bielajew & Rogers 1988). The theorem states that the response (fluence, dose, ICSD, etc.) calculated at depth in a cylindrical bin of radius $r$ by a normally incident beam of radius $R$, is equal to the response calculated at depth in a cylindrical bin of radius $R$ by a normally incident beam of radius $r$.

The geometry used to simulate the setup of the experiments performed at HIT (Dokic *et al* 2016) consisted of a 1 cm radius and 1 cm length water cylinder of resembling a water-filled well of a cell plate. The first millimeter in the proximal surface of this volume was considered for scoring as it is the position where cells were plated. The cylinder was placed behind 3.5 cm of water to reproduce the experimental setup. The radiation source was tuned to match the experimental depth dose distributions reported in (Dokic *et al* 2016).

In each run, 500,000 primary particles were simulated. The standard deviation of the dose in the bins relevant to the experiment was in the range of $6\times10^{-5}\%$ to $4\times10^{-4}\%$. Mean fluence per source particle was calculated as the path length per unit volume (Equation (15)). Mean dose per source particle was calculated as the energy deposited per unit mass.

The frequency ICSD was interpolated on the fly, based on the energy of each particle at each interaction point in the bin. That is, for each particle that laid down a track segment in a given depth bin, particle kinetic energy and type was used to retrieve

*Nanodosimetric quantities in treatment planning for charged particle therapy* 15

the corresponding $f^c(\nu)$ from our pre-calculated data described in Section 2.5. The following approach was used to interpolate $f^c(\nu)$ during the simulations. Let $E$ be the kinetic energy of particle class $c$, in units of MeV for electrons and protons, MeV/u for ions heavier than protons. For each $E$ in the interval $E_1 < E < E_2$, where $c_1$ and $c_2$ are one type of particle with sequential energies $E_1$ and $E_2$, available from our MCTS calculations, we obtained $f^c(\nu)$ by linear interpolation:

$$f^c(\nu) = f^{c_1}(\nu) + \frac{E - E_1}{E_2 - E_1} \left( f^{c_2}(\nu) - f^{c_1}(\nu) \right). \qquad (21)$$

The energy spacing was sufficiently small that logarithmic and linear interpolation gave nearly identical results. Equation (17) was then used with Equation (15) to obtain $f^{\mathscr{C}_j}(\nu)$ for the entire run. The value of $I_p^{\mathscr{C}_j}$ was then obtained using Equation (18).

Results were used to illustrate the extent to which different source ion beams with the same local fluence and $I_p^{\mathscr{C}}$ have the same biological effect. To do so, we first calculated the local fluence in the carbon ion beam used in the BEVALAC measurements at the depth of maximum dose that resulted in a dose of 2.5 Gy. The dose from this local fluence was then calculated for both sets of measurements for all of the beams at the different depths where cell survival was measured. The survival at this dose was determined from a fit to the measured survival data, using a linear-quadratic fit for the data from LBNL and a linear fit for the data from HIT. The uncertainty in survival was computed by propagating the published experimental uncertainties in alpha, $\delta\alpha$, and beta, $\delta\beta$, as follows:

$$\delta S = \sqrt{\frac{\delta S}{\delta \alpha}^2 \delta\alpha^2 + \frac{\delta S}{\delta \beta}^2 \delta\beta^2}. \qquad (22)$$

Results were then plotted against the nineteen different $I_p$ chosen for demonstration purposes. Equation (18) was used, with the fluence-weighted average frequency distribution from (Equation (17)), to calculate $N_k, k = 2, 10$ and $F_k, k = 1, 10$ at each measurement point along the depth dose curves for the distribution of particles calculated for each beam using the condensed history Monte Carlo method. The frequency distribution operators of $N_k$ and $F_k$ were those defined in Equations (3) and (4), respectively.

Finally, results from the experiments were used to demonstrate the use of the equations to show the relationship between cell survival and cluster dose. Equation (20) was used to calculate the fluence that resulted in the same $g_j^{(I_p)}$ for the value of $g_j^{(I_p)}$ corresponding to a 2.5 Gy maximum dose in the carbon beam in the BEVALAC measurements.

## 3. Results

### 3.1. Calculation of frequency ICSD

The calculated depth dose curves for the simulated monoenergetic proton, C, Ar, and Ne beams are shown in Figure 3. The cluster dose $g_j^{(N_4)}$, calculated with Equation (10),

*Nanodosimetric quantities in treatment planning for charged particle therapy*     16

is also shown. The peak to plateau ratio of the $g_j^{(N_4)}$ curves is larger than that of the depth dose curves, an indication that this $I_p$ has a stronger association with biological effect than dose.

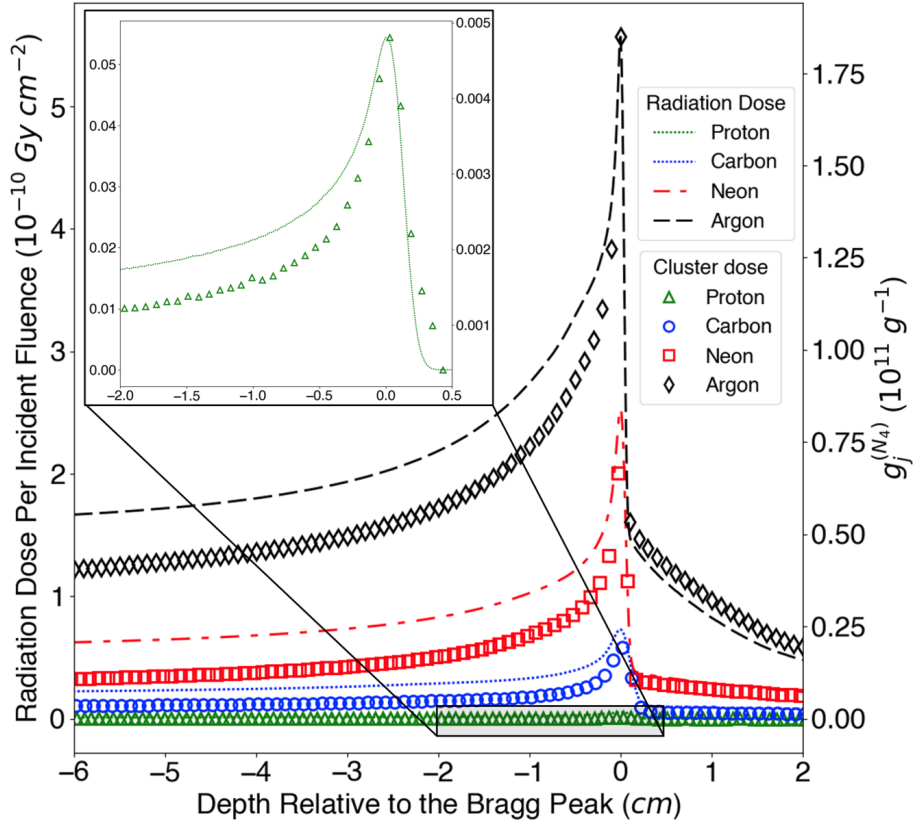

Figure 3: Radiation dose per incident fluence (lines, scale on left side of graph) and cluster dose $g^{(N_4)}$ (points, scale on right side of graph) for 100 MeV proton, 400 MeV/u C, 425 MeV/u Ne, and 570 MeV/u Ar beams. Both quantities were calculated with TOPAS in a single simulation for each beam. A magnified plot of the proton curves is shown in the inset. Zero on the distance axis corresponds to the depth of maximum dose.

The cluster dose $g_j^{(F_4)}$, calculated with Equation (11), is shown in Figure 4. We verified that the Geometry Equivalence theorem applies to $g_j^{(F_4)}$ calculation as follows. This quantity was calculated along the central axis of monoenergetic proton, C, Ar and Ne beams in the cylindrical bins of 4 cm radius for a field of 10 cm radius and also in the cylindrical bins of 10 cm radius for a field of 4 cm radius. Results, shown in Figure 4, agree within two standard deviations of calculation precision. Our use of the Geometry Equivalence theorem was critical to calculation of the frequency ICSD database in a reasonable time without loss of accuracy. The broad beam simulation would have taken 44 days on a 24 core intel server CPU. In contrast, the narrow beam calculations, converted to broad beam results with the theorem, took 7 days on the same computer.

*Nanodosimetric quantities in treatment planning for charged particle therapy* 17

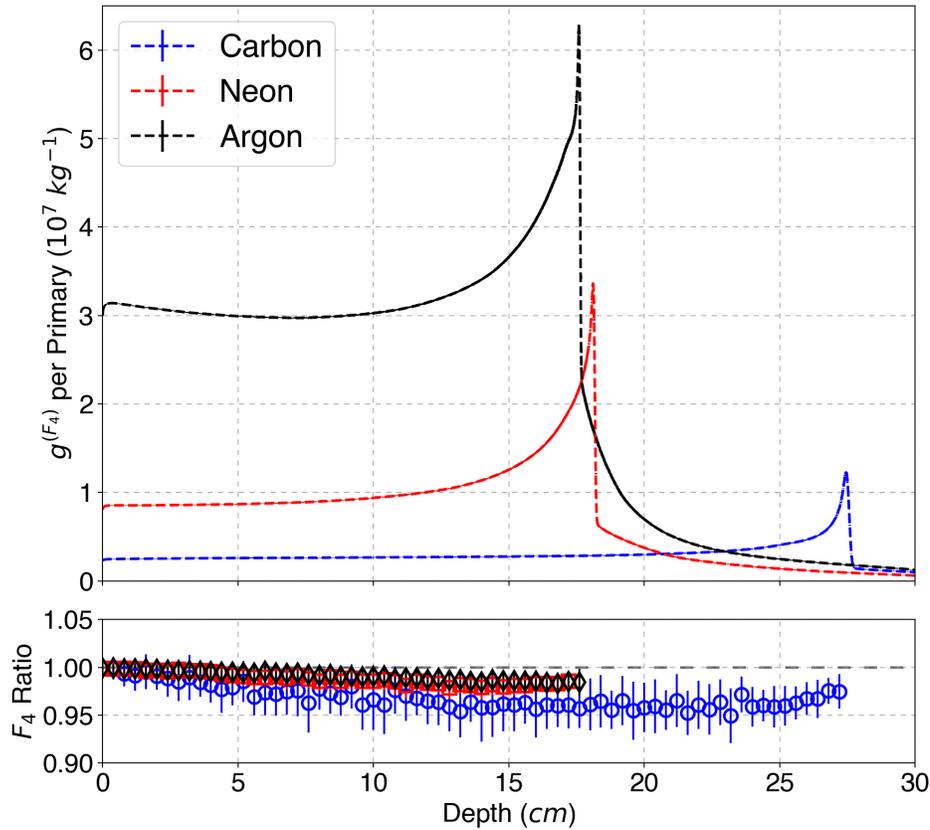

Figure 4: Upper graph: The cluster dose $g^{(F_4)}$ calculated with a 4 cm radius field in a 10 cm radius scoring bin for the ion beams in Figure 3. Lower graph: Ratio of $F_4$ calculated with a narrow (4 cm radius) field with a 10 cm radius scoring bin to that with a broad (10 cm radius) field with a 4 cm radius scoring bin for these ion beams. Error bars are 1 standard deviation calculation precision.

Frequency ICSD calculated at the depths where cell inactivation was measured (Blakely *et al* 1979), relative to the position of the maximum dose in the depth dose curve, as shown for the carbon ion beam in Figure 5.

Frequency ICSD calculated at the depth of maximum dose for the different particle types ranging from protons through argon are shown in Figure 6.

### 3.2. Association of ID parameters with cell inactivation

The relationship between measured cell survival and the $I_p$ quantities $N_k$ and $F_k$ was determined for the BEVALAC beams used in the experiment. This comparison is best done with survival measured for particles with overlapping $I_p$. The BEVALAC data used for this purpose was ideal. A particle fluence of $3.6 \times 10^6$ mm$^{-2}$ was used to calculate the dose for determining the survival. The ID parameters most closely associated with survival are those exhibiting the least separation in survival when plotted against $I_p$. For example, in Figure 7, $F_5$ is comparable to $F_6$ and preferred over the other $F_k$ shown. Plots with $k < 3$ and $k > 9$ (not shown) exhibit greater separation than those shown in

*Nanodosimetric quantities in treatment planning for charged particle therapy*　　　18

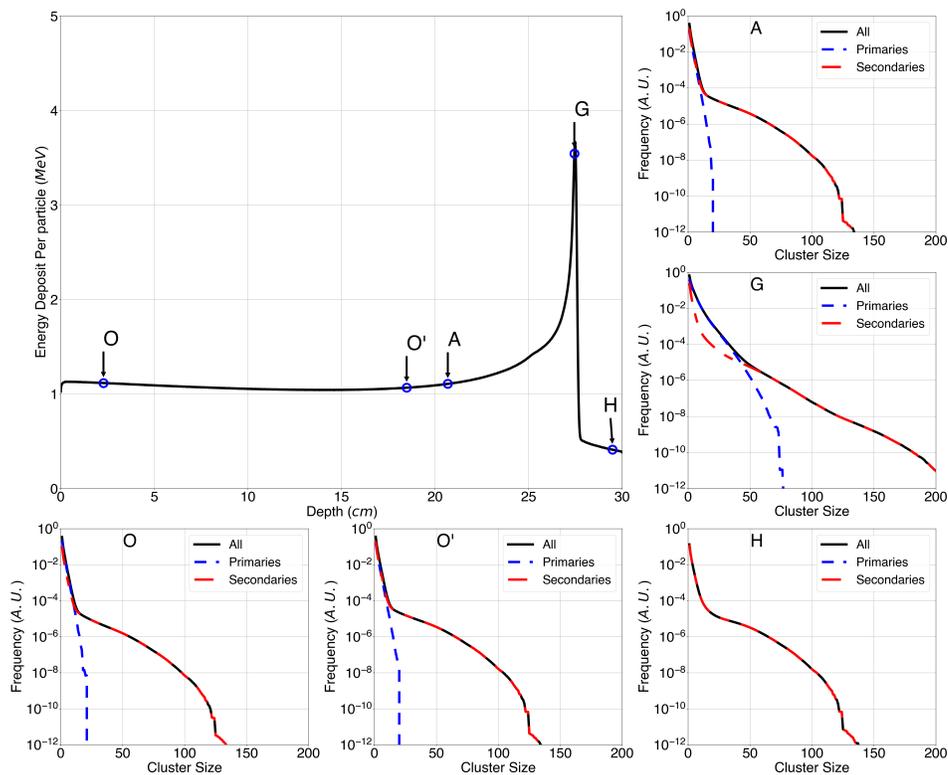

Figure 5: Central axis depth dose curve of a 400 MeV/u carbon ion beam along with five frequency ICSD calculated for primaries (dashed blue line), secondaries (solid red line) and all charged particles (dashed black line), calculated with TOPAS. The frequency ICSD, used for the calculation of $I_p$ using Equation (1), were calculated at the points indicated with arrows on the depth dose curve. The point labels match those from Figure 3 of (Blakely *et al* 1979).

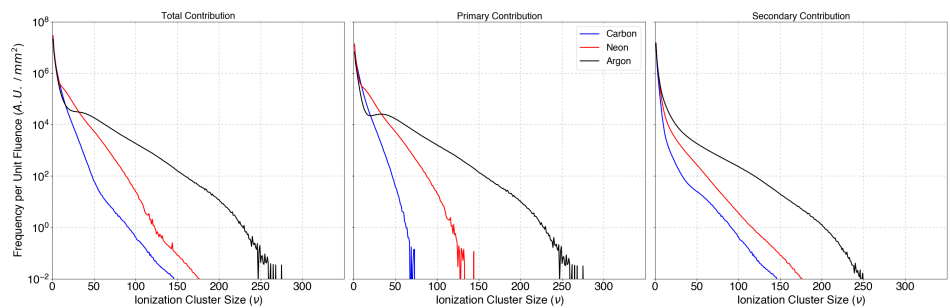

Figure 6: Frequency ICSD at the depth of maximum dose (point G in Figure 5) for the simulated 400 MeV/u C ion beam, 425 MeV/u Ne ion beam, and 570 MeV/u Ar ion beam for primaries (left), secondaries (middle) and all charged particles (right).

the figure.

The $N_k$ and $F_k$ preferred for their stronger association with cell survival for these experiments with BEVALAC beams were $N_4$ and $F_5$ for aerobic cells and $N_5$ and $F_7$ for hypoxic cells. These are plotted together, along with the association of survival with LET, in Figure 8 for aerobic cells and in Figure 9 for hypoxic cells. In these plots, survival

*Nanodosimetric quantities in treatment planning for charged particle therapy* 19

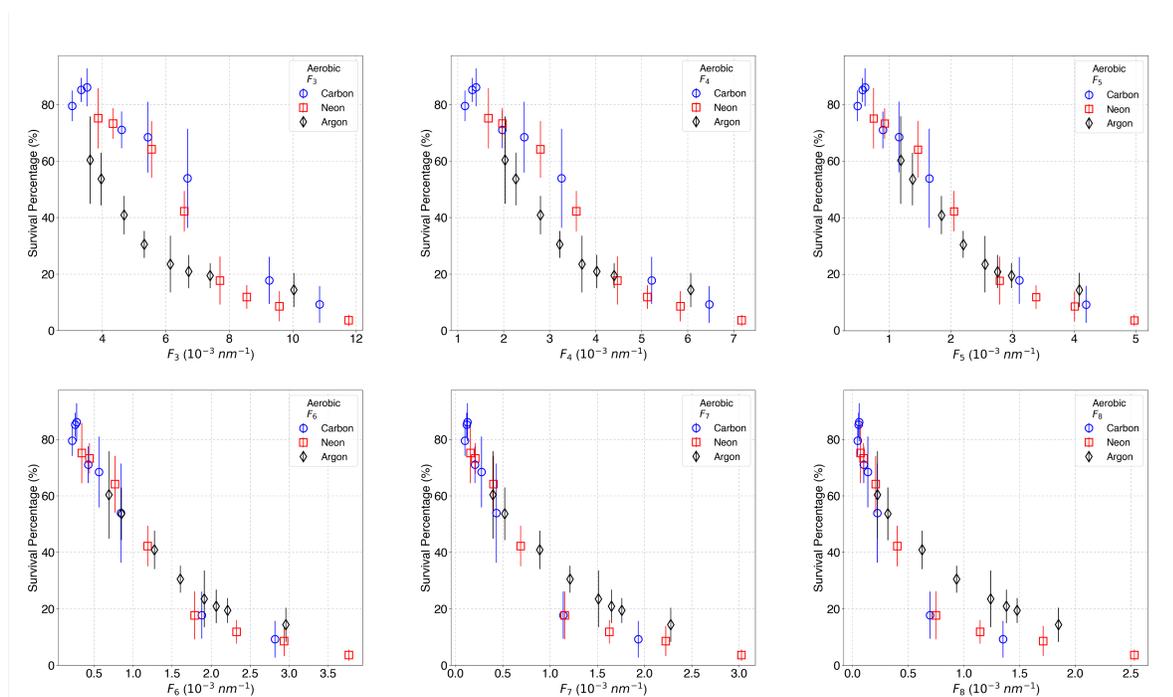

Figure 7: The dependence of the separation in cell survival measured under aerobic conditions on $F_k$, for $k = 3, ..., 8$, given a fluence of $3.6 \times 10^6$ mm$^{-2}$ for carbon (blue circles), neon (red squares) and argon (black diamonds) ions. Error bars are 1 standard deviation.

is more strongly associated with the preferred $I_p$ than with LET. Particles with an LET in the range of 100 keV/μm and above can result in a significantly different survival, outside of experimental uncertainty. Whereas survival is the same, within experimental uncertainty, for particles with the preferred $F_k$. The survival for the preferred $F_k$ is tighter than that for the preferred $N_k$.

No distinction was made in the physical properties of the ionization pattern in aerobic and hypoxic liquid water. That is, the same precalculated frequency ICSD database and thus the same resulting $I_p$ values were used for both aerobic and hypoxic conditions. This is reasonable, considering that at Normal Temperature and Pressure (NTP) oxygen constitutes only about 0.001% of the molecular composition of water (U.S.A. Environmental Protection Agency 2012).

### 3.3. Association of the cluster dose with cell inactivation

The relationship between measured cell survival and the value of each $I_p$ calculated to give the same cluster dose for the different irradiation conditions was determined for the nineteen different $I_p$. This comparison does not require particles with overlapping $I_p$. Both the HIT and BEVALAC data were used for this purpose. Results for the HIT and BEVALAC experiments are shown in separate plots since different cell lines and experimental techniques were used.

*Nanodosimetric quantities in treatment planning for charged particle therapy* 20

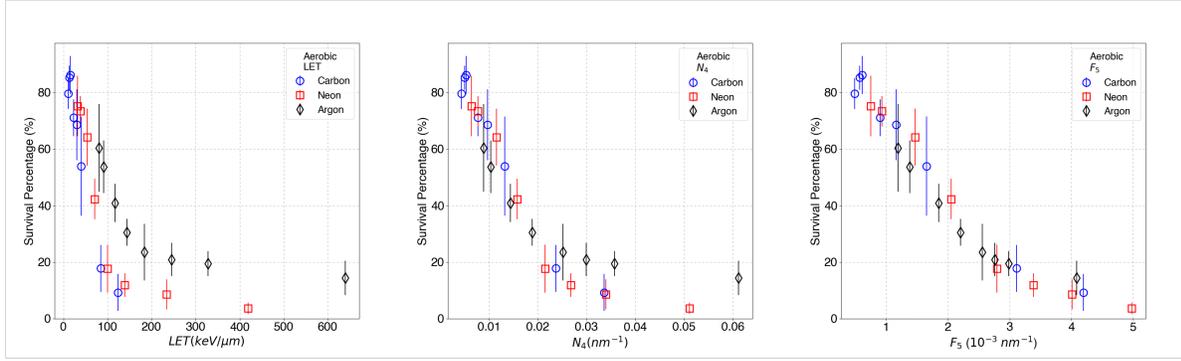

Figure 8: The association of cell survival measured under aerobic conditions with LET (left), $N_4$ (middle) and $F_5$ (right) for a fluence of $3.6 \times 10^6$ mm$^{-2}$ for carbon (blue circles), neon (red squares) and argon (black diamonds) ions. Error bars are 1 standard deviation.

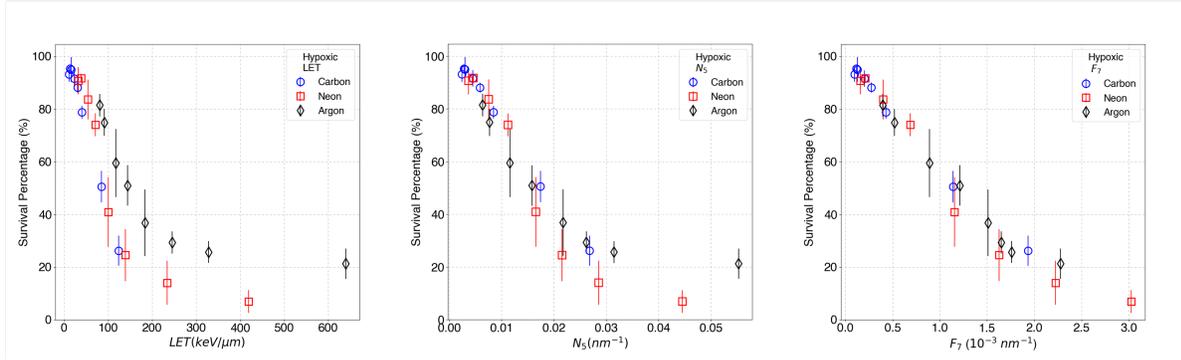

Figure 9: The association of cell survival measured under hypoxic conditions with LET (left), $N_5$ (middle) and $F_7$ (right) for a fluence of $3.6 \times 10^6$ mm$^{-2}$ for carbon (blue circles), neon (red squares) and argon (black diamonds) ions. Error bars are 1 standard deviation.

The HIT data has only four values for a given $I_p$ (one for each of the four different particles) and these do not overlap. The cluster dose was that at 2.5 Gy at the depth of maximum dose of the BEVALAC carbon beam. For $F_1$ through $F_{10}$, $g_j^{(F_k)}$ was calculated to be 212, 89.9, 40.1, 24.2, 16.1, 11.2, 7.98, 5.84, 4.33, and 3.27 $\times 10^{12}$ $g^{-1}$.

The different $I_p$ used in the demonstration have a wide range of association with the cell survival in the experiments used in the demonstration. Results for aerobic cells for $F_k, k = 4, 6$, shown in Figure 10 and Figure 11, illustrate that this type of $I_p$ has the strongest association with aerobic cell survival when $k = 5$ (plots in the middle of the figure), being the least dependent on particle type and energy for a constant value of the cluster dose $g_j^{(F_k)}$. The result applies to both types of cultured human cells considered here for demonstration purposes. It is apparent that when clusters of four or fewer ionizations in the nanoscopic scoring volume are counted, the lower $I_p$ (lower LET, higher fluence) particle classes have insufficient dose to achieve the lower survival of the higher $I_p$ (higher LET, lower fluence) particle classes. The reverse is true when

*Nanodosimetric quantities in treatment planning for charged particle therapy*      21

clusters of five or more ionizations are excluded from the count.

Results for hypoxic cells for $F_k, k = 6, 8$, shown in Figure 12, illustrate that this type of $I_p$ has the strongest association with hypoxic cell survival when $k = 7$ (plot in the middle of the figure). Considering the constant fluence plots from the previous section and the constant cluster dose plots from this section, the $I_p$ preferred for their stronger association with cell survival for the experiments considered in the demonstration were $F_5$ for aerobic cells and $F_7$ for hypoxic cells.

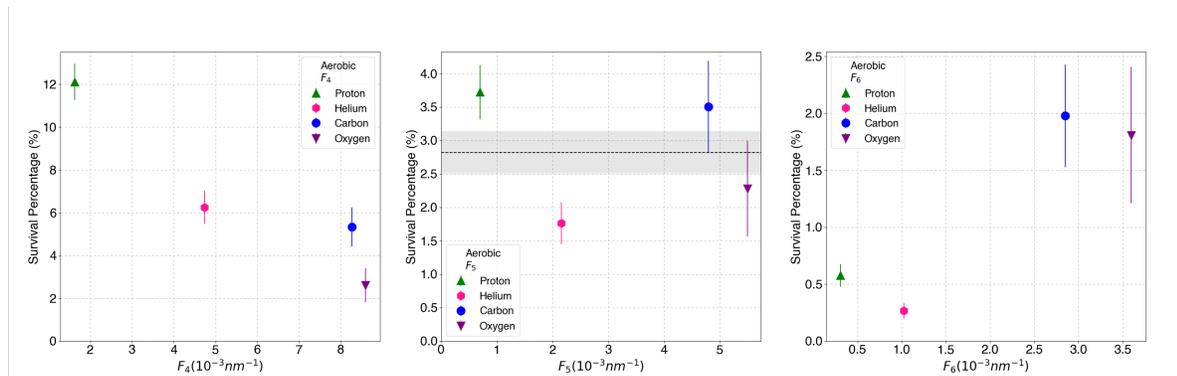

Figure 10: The relationship between aerobic cell survival and $F_4$ (left), $F_5$ (middle) and $F_6$ (right). Survival determined from measurement at the dose calculated to give the same cluster dose $g_j^{(F_k)}$ in cells irradiated at the center of the 1 cm wide SOBP for proton, helium, carbon, and oxygen beams at HIT. Error bars are 1 standard deviation, based on the published experimental uncertainty. The average survival of the points in the middle plot is shown as a dashed line, the gray region showing 1 standard deviation experimental uncertainty.

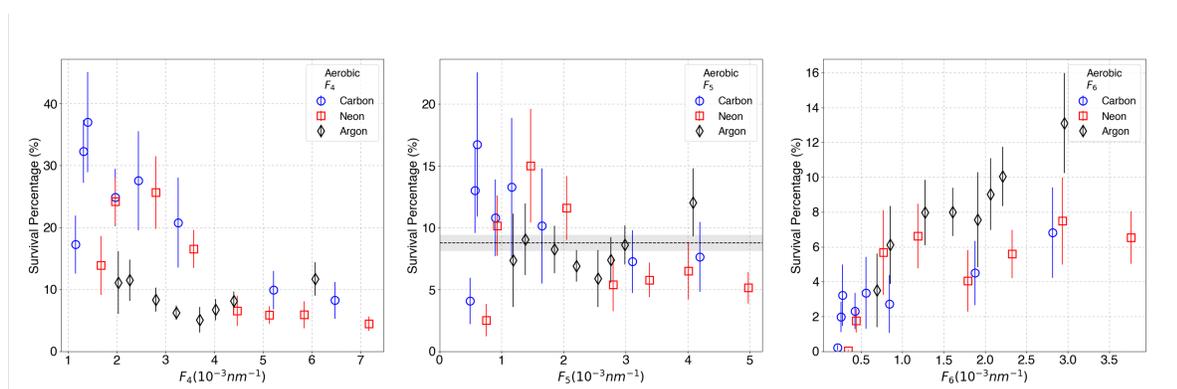

Figure 11: The relationship between aerobic cell survival and $F_4$ (left), $F_5$ (middle) and $F_6$ (right). Survival determined from measurement at the dose calculated from the particle fluence, adjusted to give the same cluster dose $g_j^{(F_k)}$ in cells irradiated at the depth of measurement in BEVALAC carbon, neon and argon beams. Error bars are 1 standard deviation, based on the published experimental uncertainty. The average survival of the points in the middle plot is shown as a dashed line, the gray region showing 1 standard deviation experimental uncertainty.

*Nanodosimetric quantities in treatment planning for charged particle therapy* 22

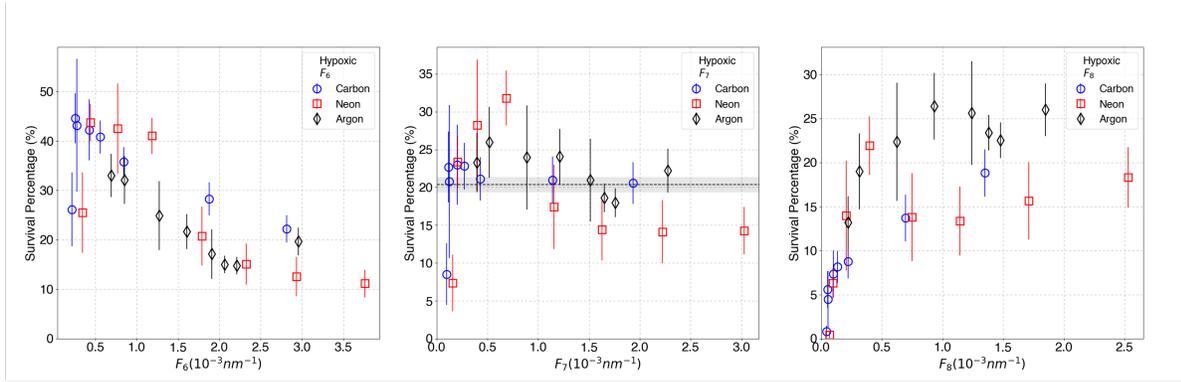

Figure 12: The relationship between hypoxic cell survival and $F_6$ (left), $F_7$ (middle) and $F_8$ (right). Survival determined from measurement at the dose calculated from the particle fluence, adjusted to give the same cluster dose $g_j^{(F_k)}$ in cells irradiated at the depth of measurement in BEVALAC carbon, neon and argon beams. Error bars are 1 standard deviation, based on the published experimental uncertainty. The average survival of the points in the middle plot is shown as a dashed line, the gray region showing 1 standard deviation experimental uncertainty.

## 4. Discussion

For charged particles, dose is the product of the local fluence with the restricted mass collision stopping power. From Equation (20), the cluster dose is the product of local fluence with $I_p^{\mathscr{C}_j}/\rho_o$, the mass ID parameter averaged over the bin. Thus $g$ is analogous to dose and $I_p$ is analogous to stopping power.

From the extensive work published in the field of nanodosimetry, it is reasonable to assert that, given a judicious choice of the operator that converts ionization detail to $I_p$ (Equation (1)), mixed ion beams with the same fluence-averaged $I_p$ (Equation (14)) would have the same biological effect in bins having the same $g_j^{(I_p)}$ (Equation (20)). In other words, we seek an $I_p$ where the biological effects are independent of the source of cluster dose.

Further, when confirmed with experimental data, we expect that we can substantially relax the requirement of a uniform fluence-averaged $I_p$. Rather, a mix of ions of different energies and fluence in each RTP voxel that provides a uniform $g_j^{(I_p)}$, within the constraints of a limited range of $I_p^{\mathscr{C}}$, would lead to uniform biological effects. For example, in a voxel irradiated by a Bragg peak of carbon ions mixed with protons, to keep $g_j^{(I_p)}$ the same, a reduced fluence of carbon ions in the Bragg peak would need to be offset by a substantially higher fluence of protons, leading to a lower $I_p^{\mathscr{C}}$. The allowed range of $I_p^{\mathscr{C}}$ will depend on the selection of $I_p$.

The possibility that preferred $I_p$ exist that would result in comparable biological effects in this scenario is supported by our results as follows. Consider Figures (10)-(12). The plot at the center of each figure shows the results for the preferred $F_k$ identified earlier; that is, $F_5$ for aerobic cells and $F_7$ for hypoxic cells. In these plots, the average survival is $2.8 \pm 0.6\%$ from the HIT data for aerobic cells, $8.8 \pm 0.3\%$ from the

*Nanodosimetric quantities in treatment planning for charged particle therapy*	23

BEVALAC data for aerobic cells, and $20.4\pm1.0\%$ from the BEVALAC data for hypoxic cells. The uncertainties quoted are 1 standard deviation, calculated with standard error propogation methods from the experimental uncertainty shown for each point. Survival of cells irradiated with the same cluster dose agrees with the average within 2 standard deviations for 79% of the 52 points. Then, aerobic cells of the same type irradiated with the same cluster dose based on $F_5$, shown in Figures (10) and (11), and hypoxic cells irradiated with the same cluster dose based on $F_7$, shown in Figure 12, had a comparable survival. The results apply to both clinical scanned beams, from protons through oxygen (Dokic *et al* 2016), and scattered pristine beams, from carbon through argon (Blakely *et al* 1979), as well as different types of cells and cells irradiated under both aerobic and hypoxic conditions.

It is our hope that preferred $I_p$ exist that allow ID-based RTP on $g_j^{(I_p)}$ alone. We believe this to be plausible, although it has yet to be determined.

### 4.1. Model choices

We made choices in the ID model with the objective of presenting a model amendable to ID-based RTP. These choices are neither modeling assumptions nor assertions that need to be proven. Of importance to us was to devise an algorithm with which the ID-based RTP inverse problem, posed by the model, will be solved, if applied to a situation where the modeling assumptions hold. In that respect, one would prefer a model which represents the real-world with less fidelity but for which a tractable algorithm for ID-based RTP exists, or can be devised, over a model which is more sophisticated and/or accurate but for which no solution method for the ID-based RTP problem is available.

Regarding the details of the MCTS simulation used to calculate the frequency ICSD, the choice of source, encompassing volume, and target size, shape and distribution, for example, may significantly alter the result. This opens the possibility of preferred choices in the details of how the frequency ICSD is determined, in terms of the impact of these choices on the association of the resulting $I_p$ with biological effects.

Further, the ID, mathematically described by $f^c(\nu)$, was embedded in an ID parameter $I_p^c$ (Equation (14)). This simplifies the problem of bridging the large gap in scale between nanodosimetric quantities and RTP in macroscopic voxels. The operator $G_p$ was assumed to be linear. This opens up the possibility for using an existing algorithm or devising an algorithm for the ID-based RTP problem using $I_p^{\mathscr{C}_j}$ (Equation (14) or (18)) and/or $g_j^{(I_p)}$ (Equation (5) or (10)). The possibility of applying a nonlinear operator remains open. This could expand our model by generalizing the relationships derived for a linear operator in (Equations (7)-(11)).

### 4.2. Calculation of frequency ICSD and ID parameters

We have demonstrated the calculation of frequency ICSD for ions ranging from protons to argon. We have also demonstrated the calculation of $I_p$ required by the ID model for the examples. Our results provide evidence of the value of doing such calculations.

The configuration of the source and geometry used in the simulation of the particle tracks was chosen by way of example and is not intended to represent a final or best choice. The results are for one of many possible configurations. The calculated quantities may depend on the configuration, such as the target sensitive volume size and shape. Other configurations may lead to a closer association between $I_p$ and biological effects.

*4.3. A demonstration of the selection of preferred ID parameters*

The quantities LET, $N_k, k = 2,3,...10$ ($N_1 = N_2$), and $F_k, k = 1,2,...10$, were compared to measured cell survival for cells irradiated with carbon, neon and argon ions from a BEVALAC under both aerobic and hypoxic conditions. The closest association between the $I_p$ and survival occurred for $N_4$ and $F_5$ under aerobic conditions and $N_5$ and $F_7$ under hypoxic conditions. Results for select $I_p$ are shown in Figure 8 for aerobic cells and Figure 9 for hypoxic cells. There were significant differences in measured cell survival for particles having the same LET or the same $N_k$, including $N_4$ for well oxygenated cells and $N_5$ for hypoxic cells, that show the closest relationship to survival. Of the nineteen $I_p$ considered in the demonstration, the preferred $I_p$ were $F_5$ for aerobic cells and $F_7$ for hypoxic cells. These $I_p$ exhibited a close association with survival, within experimental uncertainty. This encouraging result demonstrates that there exist preferred $I_p$ that are more closely associated with biological effects than LET.

The particle fluence for plotting survival against LET and the selected $I_p$ was chosen to result in a maximum dose of 2.5 Gy in the carbon beam from the BEVALAC. This dose was chosen since a similar dose is commonly used for a single radiotherapy fraction in carbon therapy. The resulting fluence and cluster dose were appropriate to obtain survival for all of the particles and depths of measurement without extrapolation of the measured survival curves. Since dose-response curves are close to linear when plotting the logarithm of cell survival against dose, a different fluence would show a comparable association of $I_p$ with cell survival.

The examples demonstrate how to use the formalism to experimentally determine preferred $I_p$ for particle therapy RTP. In our demonstration of $I_p$ evaluation, we chose different cell types irradiated under aerobic conditions with different types of beams, and with heavier ion beams, we chose data with cells irradiated under both aerobic and hypoxic conditions. This allowed us to show how well different $N_k$ and $F_k$ associate with cell survival under different conditions. In fact, $F_5$ was associated with cell inactivation within experimental uncertainty for cells irradiated under aerobic conditions, for two different types of cells and two different beam delivery systems, $F_7$ for the same type of cells irradiated under hypoxic conditions. The larger cluster sizes associated with comparable cell survival in hypoxic cells can be explained by the lower probability of an ionization to be converted to a DNA strand break compared to aerobic cells (Schulte *et al* 2003). The result that $F_5$ was more closely associated with survival for aerobic cells, $F_7$ for hypoxic cells, could be dealt with in patient treatment by mapping oxygen levels in the tumor then calculating the cluster dose with the $F_k$ appropriate for the aerobic

*Nanodosimetric quantities in treatment planning for charged particle therapy*    25

and hypoxic regions.

It is beyond the scope of this paper to evaluate the many possible choices for $I_p$. An accurate determination of preferred $I_p$ requires a much larger set of measurements covering a wide range of cell types and biological endpoints. This includes the choice of whether to use the ICSD or an alternative frequency distribution to characterize the ID, consideration of specifics in the calculation of the frequency distribution (in the case of ICSD one can choose different source distributions, target volumes, etc.), and selection of the operator to convert the frequency distribution to $I_p$. It also includes determination of how well the selected $I_p$ associate with biological effects, ideally covering a wide range of parameters such as biological materials (different cell types, different states of growth, different tissues and organs in in-vivo pre-clinical studies, etc.), environment (oxygen tension, etc.), and other compounding factors, a most ambitious project. We believe that our examples provide compelling evidence for the community to embark on such a venture.

### 4.4. A demonstration of the relationship between biological effect and cluster dose

The data from the cell survival experiments used to demonstrate selection of preferred $I_p$, discussed in the previous selection, was further used to demonstrate the relationship between cell survival and cluster dose. This data was augmented with aerobic cell survival experiments performed with clinical beams of proton, helium, carbon and oxygen at HIT, to extend these results down to protons. There was a close relationship between aerobic cell survival and $F_5$ as well as between hypoxic cell survival and $F_7$. This demonstrates the existence of $I_p$ that have high potential for use in particle beam treatment planning.

### 4.5. Application to ID-based RTP

The ID model presented here provides a path to clinical translation to intensity modulated particle therapy. A conservative approach would be to first introduce ID-based RTP into the clinic using simultaneous optimization of currently used RBE-weighted dose along with $I_p$ that are found to be closely associated with biological effects, along the lines demonstrated by (Burigo *et al* 2019).

Based on previously published work (Burigo *et al* 2019), we can formulate a constrained optimization problem as

$$\vec{w}^* = \arg\min \chi(\vec{w}) := \sum_{n=1}^{N} (p_{n,D} f_{n,D}(\vec{w}) + p_{n,I_p^\mathscr{C}} f_{n,I_p^\mathscr{C}}(\vec{w}) + p_{n,g} f_{n,g}(\vec{w})) \quad (23)$$
$$\text{such that } ((I_p^\mathscr{C})^{\min})_j \leq (I_p^\mathscr{C}(\vec{w}))_j \leq ((I_p^\mathscr{C})^{\max})_j, \text{ for } j = 1, 2, \ldots, M,$$
$$\text{and } (g^{\min})_j \leq (g(\vec{w}))_j \leq (g^{\max})_j, \text{ for } j = 1, 2, \ldots, M,$$

where $\vec{w}$ is the incoming fluence represented as a vector of pencil beam weights

*Nanodosimetric quantities in treatment planning for charged particle therapy*     26

(intensities), the objective functions $f_{n,D}(\vec{w})$, $f_{n,I_p^{\mathscr{C}}}(\vec{w})$ and $f_{n,g}(\vec{w})$ are based on (RBE-weighted-) dose, on $I_p^{\mathscr{C}}$, and on cluster dose, respectively, within each planning structure $n$, with a totality of $N$ structures.

The objective functions are weighted by, so called, "penalty factors" $p_{n,D}$ (for RBE), $p_{n,I_p^{\mathscr{C}}}$ and $p_{n,g}$ (for ID), chosen by the planner to handle the trade-off between (RBE-weighted) dose and ID-based optimization. The ratios of $p_{n,I_p^{\mathscr{C}}}$ and $p_{n,g}$ to $p_{n,D}$ also represent the confidence in the ID versus RBE formalism for each structure $n$.

The second line of Equation (23) represents the constraints that use the prescribed bounds $((I_p^{\mathscr{C}})^{\min})_j \leq (I_p^{\mathscr{C}}(\vec{w}))_j \leq ((I_p^{\mathscr{C}})^{\max})_j$ on the ID parameter, for each voxel $j$. The third line of Equation (23) represents the constraints that use the prescribed bounds $(g^{\min})_j \leq (g(\vec{w}))_j \leq (g^{\max})_j$ on the cluster dose, for each voxel $j$.

Pencil beam weights $\vec{w}^*$ would be chosen using conventional optimization methods used in RTP or alternative approaches like the superiorization methodology, (Herman *et al* 2012, Censor 2022, Censor 2023) In a target volume, pencil beam weights $\vec{w}^*$ that result from solving the optimization problem of Equation (23) should be such that they would yield a uniform $I_p^{\mathscr{C}}$. This can be achieved by defining suitable objective functions $f_{n,I_p^{\mathscr{C}}}$, e.g., functions that are based on the variance of $I_p^{\mathscr{C}}$ over all voxels, together with the common $f_{n,D}(\vec{w})$. In normal tissues, hard constraints bounds $((I_p^{\mathscr{C}})^{\max})_j$ would be defined to keep the corresponding $(I_p^{\mathscr{C}}(\vec{w}))_j$ below reasonable limits. The same considerations apply to the constraints on $g_j^{(I_p)}$.

The constraints would best be developed by the research and clinical community based on experimental and clinical evidence. We anticipate the use of both published measurements and experiments designed explicitly to help determine reasonable bounds for the constraints. This will be complemented by knowledge gained through experience of the range of $I_p$ and $g_j^{(I_p)}$ observed in clinical treatments.

The RTP problem in Equation (23) is formulated using both $I_p^{\mathscr{C}}$ and $g_j^{(I_p)}$. Instead, the problem can be formulated without using $I_p^{\mathscr{C}}$, facilitating simultaneous optimization of radiation dose and cluster dose in RTP. Alternatively, the optimization could be done with cluster dose alone. Before this can be assessed, we first need to determine a set of preferred $I_p$. Once this is available, we will be able to work on determining which planning approach to use in the clinic.

This suggests the following path to clinical translation of ID-based RTP. After gaining experimental and clinical experience with the simultaneous optimization approach, ID-based RTP could be done alone, applying the objective functions on uniformity of the ID parameter and cluster dose in the target volume and hard maximum constraints on these in critical structures. The next step would be to optimize the cluster dose alone.

## 5. Conclusions

We have defined ID parameter $I_p$, a collapsed representation of the detailed spatial distribution of ionizations along particle tracks. We have also defined the cluster dose $g^{(I_p)}$, the product of fluence with the mass $I_p$, analogous to the relationship between radiation dose and the restricted mass collision stopping power. Together these quantities encapsulate the detailed stochastic distribution of ionization from particle tracks into physical quantities for association with biological effects for their practical application in treatment planning.

The mathematical model of ID presented here sets the stage for determining the degree of association between biological effects and details of the distribution of ionization along tracks of charged particles of different types and energies. The degree of association is clearly impacted by the choice of $I_p$.

It is reasonable to assert that there exists one or more different $I_p$ that are more closely associated with biological effects than LET and current microdosimetric RBE-based models used in particle RTP, although results were not shown in this paper to directly support the latter. We base this assertion on the growing literature in nanodosimetry, such as the known strong association of ionization cluster complexity with biological effects, supported by the results shown in our examples.

An accurate determination of preferred $I_p$ requires a much larger set of measurements than we presented for demonstration purposes. There is a great deal of published data for various particles allowing the calculation of a wide range of $I_p$ values and with a variety of biological endpoints that could be evaluated to determine preferred $I_p$ that are best associated with biological or clinical effects.

Our mathematical model provides a practical means to employ our knowledge of the physics and radiobiology of ionization along particle tracks obtained at the nanometer scale to RTP at the millimeter scale for patients undergoing particle radiotherapy. Our work may provide a path to clinical translation of ID-based RTP, with the potential to use cluster dose in place of radiation dose.

**Acknowledgments**

We are grateful to Dr. Ziad Francis of Saint Joseph University of Beirut, RU Mathematics and Modelling, Beirut, Lebanon for providing ionization cross-sectional data and to Tomasz Walenta for the preparation of the graphical illustration of Figure 1. EAB acknowledges support from NASA grant #80JSC021T0017 (NNJ16HP22I) under contract #DE-AC02-05CH11231 with the U.S. Department of Energy, Office of Science. The work of Yair Censor was supported by the ISF-NSFC joint research plan Grant Number 2874/19. The work was partially supported by NIH grants P20 CA183640 and R01CA266467.

## References

Bielajew A F & Rogers D W O 1988 *in* T. M Jenkins, W. R Nelson & A Rindi, eds, 'Monte Carlo Transport of Electrons and Photons' Plenum Press pp. 407–419.

Blakely E A 1992 Cell inactivation by heavy charged particles *Radiation and Environmental Biophysics* **31**(3), 181–196.
**URL:** https://doi.org/10.1007/BF01214826

Blakely E A, Tobias C A, Yang T C, Smith K C & Lyman J T 1979 Inactivation of human kidney cells by high-energy monoenergetic heavy-ion beams *Radiation Research* **80**(1), 122–160.
**URL:** https://doi.org/10.2307/3575121

Bueno M, Schulte R, Meylan S & Villagrasa C 2015 Influence of the geometrical detail in the description of DNA and the scoring method of ionization clustering on nanodosimetric parameters of track structure: A Monte Carlo study using Geant4-DNA *Physics in Medicine and Biology* **60**(21), 8583–8599.
**URL:** https://doi.org/10.1088/0031-9155/60/21/8583

Burigo L N, Ramos-Méndez J, Bangert M, Schulte R W & Faddegon B 2019 Simultaneous optimization of RBE-weighted dose and nanometric ionization distributions in treatment planning with carbon ions *Physics in Medicine and Biology* **64**(1), 15015.
**URL:** http://dx.doi.org/10.1088/1361-6560/aaf400

Casiraghi M & Schulte R W 2015 Nanodosimetry-based plan optimization for particle therapy *Computational and Mathematical Methods in Medicine* **2015**, 908971.
**URL:** https://doi.org/10.1155/2015/908971

Censor Y 2022 'Superiorization and Perturbation Resilience of Algorithms: A Bibliography compiled and continuously updated'.
**URL:** http://math.haifa.ac.il/yair/bib-superiorization-censor.html

Censor Y 2023 Superiorization: The asymmetric roles of feasibility-seeking and objective function reduction *Applied Set-Valued Analysis and Optimization, Available on arXiv at: https://arxiv.org/abs/2212.13182.* **In press**.

Charlton D E, Nikjoo H & Humm J L 1989 Calculation of Initial Yields of Single- and Double-strand Breaks in Cell Nuclei from Electrons, Protons and Alpha Particles *International Journal of Radiation Biology* **56**(1), 1–19.
**URL:** https://doi.org/10.1080/09553008914551141

Chatzipapas K P, Papadimitroulas P, Emfietzoglou D, Kalospyros S A, Hada M, Georgakilas A G & Kagadis G C 2020 Ionizing radiation and complex DNA damage: Quantifying the radiobiological damage using Monte Carlo simulations *Cancers* **12**(4), 1–23.
**URL:** https://doi.org/10.3390/cancers12040799

Combs S, Ellerbrock M, Haberer T, Habermehl D, Hoess A, Jäkel O, Jensen A, Klemm S, Münter M, Naumann J, Nikoghosyan A, Oertel S, Parodi K, Rieken S & Debus J 2010 Heidelberg Ion Therapy Center (HIT): Initial clinical experience in the first 80 patients *Acta Oncol.* **49**(7), 1132–1140.

Conte V, Bianchi A & Selva A 2023 Track Structure of Light Ions: The Link to Radiobiology *International Journal of Molecular Science* **24**(6), 5826.

Conte V, Selva A, Colautti P, Hilgers G & Rabus H 2017 Track structure characterization and its link to radiobiology *Radiation Measurements* **106**, 506–511.
**URL:** https://doi.org/10.1016/j.radmeas.2017.06.010

Conte V, Selva A, Colautti P, Hilgers G, Rabus H, Bantsar A, Pietrzak M & Pszona S 2018 Nanodosimetry: Towards a new concept of radiation quality *Radiation Protection Dosimetry* **180**(1-4), 150–156.
**URL:** http://dx.doi.org/10.1093/RPD/NCX175

Curtis S 1974 Plans for the high-energy, heavy-ion facility (BEVALAC) at Berkeley *European Journal of Cancer* **10**(6), 388.

Deng W, Yang Y, Liu C, Bues M, Mohan R, Wong W W, Foote R H, Patel S H & Liu W 2021 A critical review of LET-based intensity-modulated proton therapy plan evaluation and optimization for head and neck cancer management *International Journal of Particle Therapy* **8**(1), 36–49.
  **URL:** http://dx.doi.org/10.14338/IJPT-20-00049.1

Dokic I, Mairani A, Niklas M, Zimmermann F, Chaudhri N, Krunic D, Tessonnier T, Ferrari A, Parodi K, Jäkel O, Debus J, Haberer T & Abdollahi A 2016 Next generation multi-scale biophysical characterization of high precision cancer particle radiotherapy using clinical proton, helium-, carbon- and oxygen ion beams *Oncotarget* **7**(35), 56676–56689.

Ebner D K, Frank S J, Inaniwa T, Yamada S & Shirai T 2021 The Emerging Potential of Multi-Ion Radiotherapy *Frontiers in Oncology* **11**.
  **URL:** https://www.frontiersin.org/article/10.3389/fonc.2021.624786

Faddegon B, Ramos-Méndez J, Schuemann J, McNamara A, Shin J, Perl J & Paganetti H 2020 The TOPAS tool for particle simulation, a Monte Carlo simulation tool for physics, biology and clinical research. *Physica Medica, European Journal of Medical Physics* **72**, 114–121.
  **URL:** http://dx.doi.org/10.1016/j.ejmp.2020.03.019

Falk M & Hausmann M 2020 A paradigm revolution or just better resolution-will newly emerging superresolution techniques identify chromatin architecture as a key factor in radiation-induced DNA damage and repair regulation? *Cancers* **13**(1), 18.
  **URL:** http://dx.doi.org/10.3390/cancers13010018

Fossati P, Matsufuji N, Kamada T & Karger C P 2018 Radiobiological issues in prospective carbon ion therapy trials. *Medical Physics* **45**(11), e1096–e1110.
  **URL:** http://dx.doi.org/10.1002/mp.12506

Goodhead D T 1994 Initial events in the cellular effects of ionizing radiations: clustered damage in DNA *International Journal of Radiation Biology* **65**(1), 7–17.
  **URL:** https://doi.org/10.1080/09553009414550021

Grosswendt B, Pszona S & Bantsar A 2007 New descriptors of radiation quality based on nanodosimetry, a first approach *Radiation Protection Dosimetry* **126**(1-4), 432–444.
  **URL:** http://dx.doi.org/10.1093/rpd/ncm088

Hagiwara Y, Oike T, Niimi A, Yamauchi M, Sato H, Limsirichaikul S, Held K D, Nakano T & Shibata A 2019 Clustered DNA double-strand break formation and the repair pathway following heavy-ion irradiation *Journal of Radiation Research* **60**(1), 69–79.
  **URL:** https://doi.org/10.1093/jrr/rry096

Herman G T, Garduno E, Davidi R & Censor Y 2012 Superiorization: An optimization heuristic for medical physics. *Medical Physics* **39**(9), 5532–5546.
  **URL:** http://dx.doi.org/10.1118/1.4745566

Inaniwa T, Furukawa T, Kase Y, Matsufuji N, Toshito T, Matsumoto Y, Furusawa Y & Noda K 2010 Treatment planning for a scanned carbon beam with a modified microdosimetric kinetic model *Phys. Med. Biol.* **55**(22), 6721–6737.

Incerti S, Kyriakou I, Bernal M A, Bordage M C, Francis Z, Guatelli S, Ivanchenko V, Karamitros M, Lampe N, Lee S B, Meylan S, Min C H, Shin W G, Nieminen P, Sakata D, Tang N, Villagrasa C, Tran H N & Brown J M C 2018 Geant4-DNA example applications for track structure simulations in liquid water: A report from the Geant4-DNA project *Medical Physics* **45**(8), e722–e739.
  **URL:** http://dx.doi.org/10.1002/mp.13048

Jäkel O, Kraft G & Karger C 2022 The history of ion beam therapy in Germany *Z Med Phys* **32**(1), 6–22.

Karger C P & Peschke P 2018 RBE and related modeling in carbon-ion therapy *Physics in Medicine and Biology* **63**(1), 01TR02.
  **URL:** http://dx.doi.org/10.1088/1361-6560/aa9102

Kopp B, Mein S, Dokic I, Harrabi S, Böhlen T T, Haberer T, Debus J, Abdollahi A & Mairani A 2020 Development and validation of single field multi-ion particle therapy treatments. *International Journal of Radiation Oncology, Biology, Physics* **106**(1), 194–205.

**URL:** http://dx.doi.org/10.1016/j.ijrobp.2019.10.008

Lazar A A, Schulte R, Faddegon B, Blakely E A & Roach M r 2018 Clinical trials involving carbon-ion radiation therapy and the path forward. *Cancer* **124**(23), 4467–4476.
**URL:** http://dx.doi.org/10.1002/cncr.31662

Lühr A, von Neubeck C, Krause M & Troost E G C 2018 Relative biological effectiveness in proton beam therapy - Current knowledge and future challenges. *Clinical and Translational Radiation Oncology* **9**, 35–41.
**URL:** http://dx.doi.org/10.1016/j.ctro.2018.01.006

McNamara A, Willers H & Paganetti H 2019 Modelling variable proton relative biological effectiveness for treatment planning *The British Journal of Radiology* **92**, 20190334.
**URL:** https://doi.org/10.1259/bjr.20190334

Michalik V 1993 Energy Deposition Clusters in Nanometer Regions of Charged-Particle Tracks *Radiation Research* **134**(3), 265–270.

Nikitaki Z, Pariset E, Sudar D, Costes S V & Georgakilas A G 2020 In situ detection of complex DNA damage using microscopy: A rough road ahead *Cancers* **12**(11), 3288.
**URL:** http://dx.doi.org/10.3390/cancers12113288

Nystrom H, Jensen M F & Nystrom P W 2020 Treatment planning for proton therapy: What is needed in the next 10 years? *The British Journal of Radiology* **93**(1107), 20190304.
**URL:** http://dx.doi.org/10.1259/bjr.20190304

Paganetti H 2014 Relative biological effectiveness values for proton beam therapy. Variations as a function of biological endpoint, dose, and linear energy transfer *Physics in Medicine and Biology* **59**(22), 419–72.
**URL:** http://dx.doi.org/10.1088/0031-9155/59/22/R419

Palmans H, Rabus H, Belchior A L, Bug M U, Galer S, Giesen U, Gonon G, Gruel G, Hilgers G, Moro D, Nettelbeck H, Pinto M, Pola A, Pszona S, Schettino G, Sharpe P H, Teles P, Villagrasa C & Wilkens J J 2015 Future development of biologically relevant dosimetry *The British Journal of Radiology* **88**, 20140392.
**URL:** http://dx.doi.org/10.1259/bjr.20140392

Perl J, Shin J, Schumann J, Faddegon B & Paganetti H 2012 TOPAS: An innovative proton Monte Carlo platform for research and clinical applications *Medical Physics* **39**(11), 6818–6837.
**URL:** http://dx.doi.org/10.1118/1.4758060

Pietrzak M, Mietelska M, Bancer A, Rucinski A & Brzozowska B 2021 Geant4-DNA modeling of nanodosimetric quantities in the Jet Counter for alpha particles. *Physics in Medicine and Biology* **66**(22), 225008.
**URL:** http://dx.doi.org/10.1088/1361-6560/ac33eb

Pinto M, Prise K M & Michael B D 2002 Double strand break rejoining after irradiation of human fibroblasts with X rays or alpha particles: PFGE studies and numerical models. *Radiation Protection Dosimetry* **99**(1-4), 133–136.
**URL:** http://dx.doi.org/10.1093/oxfordjournals.rpd.a006743

Rabus H, Ngcezu S A, Braunroth T & Nettelbeck H 2020 "Broadscale" nanodosimetry: Nanodosimetric track structure quantities increase at distal edge of spread-out proton Bragg peaks *Radiation Physics and Chemistry* **166**(July 2019), 108515.
**URL:** https://doi.org/10.1016/j.radphyschem.2019.108515

Ramos-Méndez J, Burigo L N, Schulte R, Chuang C & Faddegon B 2018 Fast calculation of nanodosimetric quantities in treatment planning of proton and ion therapy *Physics in Medicine and Biology* **63**(23), 235015.
**URL:** http://dx.doi.org/10.1088/1361-6560/aaeeee

Roach M r, Schulte R, Mishra K, Faddegon B, Barani I, Lazar A & Blakely E A 2016 New clinical and research programs in particle beam radiation therapy: The University of California San Francisco perspective *International Journal of Particle Therapy* **2**(3), 471–473.
**URL:** http://dx.doi.org/10.14338/IJPT-15-00025.1

Roobol S J, van den Bent I, van Cappellen W A, Abraham T E, Paul M W, Kanaar R, Houtsmuller A B, van Gent D C & Essers J 2020 Comparison of high-and low-let radiation-induced DNA double-strand break processing in living cells *International Journal of Molecular Sciences* **21**(18), 6602.
  **URL:** http://dx.doi.org/10.3390/ijms21186602

Rucinski A, Biernacka A & Schulte R 2021 Applications of nanodosimetry in particle therapy planning and beyond *Physics in Medicine and Biology* **66**(24), 24TR01.
  **URL:** https://doi.org/10.1088/1361-6560/ac35f1

Scholz M, Kellerer A M, Kraft-Weyrather W & Kraft G 1997 Computation of cell survival in heavy ion beams for therapy. The model and its approximation *Radiat Environ Biophys* **36**(1), 59–66.

Schuemann J, McNamara A L, Ramos-Méndez J, Perl J, Held K D, Paganetti H, Incerti S & Faddegon B 2019 TOPAS-nBio: An extension to the TOPAS simulation toolkit for cellular and sub-cellular radiobiology *Radiation Research* **191**(2), 125–138.
  **URL:** http://dx.doi.org/10.1667/rr15226.1

Schuemann J, McNamara A L, Warmenhoven J W, Henthorn N T, Kirkby K J, Merchant M J, Ingram S, Paganetti H, Held K D, Ramos-Mendez J, Faddegon B, Perl J, Goodhead D T, Plante I, Rabus H, Nettelbeck H, Friedland W, Kundrát P, Ottolenghi A, Baiocco G, Barbieri S, Dingfelder M, Incerti S, Villagrasa C, Bueno M, Bernal M A, Guatelli S, Sakata D, Brown J M C, Francis Z, Kyriakou I, Lampe N, Ballarini F, Carante M P, Davídková M, Štěpán V, Jia X, Cucinotta F A, Schulte R, Stewart R D, Carlson D J, Galer S, Kuncic Z, Lacombe S, Milligan J, Cho S H, Sawakuchi G, Inaniwa T, Sato T, Li W, Solov'yov A V, Surdutovich E, Durante M, Prise K M & McMahon S J 2018 A new standard DNA damage (SDD) data format *Radiation Research* **191**(1), 76–92.
  **URL:** http://dx.doi.org/10.1667/rr15209.1

Schulte R, Bashkirov V, Garty G, Leloup C, Shchemelinin S, Breskin A, Chechik R, Milligan J & Grosswendt B 2003 Ion-counting nanodosimetry: Current status and future applications *Australasian Physical and Engineering Sciences in Medicine* **26**(4), 149–155.

Stasica P, Baran J, Granja C, Krah N, Korcyl G, Oancea C, Pawlik-Niedźwiecka M, Niedźwiecki S, Rydygier M, Schiavi A, Rucinski A & Gajewski J 2020 A Simple Approach for Experimental Characterization and Validation of Proton Pencil Beam Profiles *Frontiers in Physics* **8**, 346.
  **URL:** http://dx.doi.org/10.3389/fphy.2020.00346

U.S.A. Environmental Protection Agency 2012 'Dissolved Oxygen and Biochemical Oxygen Demand'.
  **URL:** https://archive.epa.gov/water/archive/web/html/vms52.html

Vignard J, Mirey G & Salles B 2013 Ionizing-radiation induced DNA double-strand breaks: a direct and indirect lighting up. *Radiotherapy and Oncology* **108**(3), 362–369.
  **URL:** http://dx.doi.org/10.1016/j.radonc.2013.06.013

Wozny A S & Rodriguez-Lafrasse C 2023 The 'stealth-bomber' paradigm for deciphering the tumour response to carbon-ion irradiation *British Journal of Cancer* **128**, 1429–1438.

Yang Y, Vargas C E, Bhangoo R S, Wong W W, Schild S E, Daniels T B, Keole S R, Rwigema J C M, Glass J L, Shen J, DeWees T A, Liu T, Bues M, Fatyga M & Liu W 2021 Exploratory investigation of dose-linear energy transfer (LET) volume histogram (DLVH) for adverse events study in intensity modulated proton therapy (IMPT). *International Journal of Radiation Oncology, Biology, Physics* **110**(4), 1189–1199.
  **URL:** http://dx.doi.org/10.1016/j.ijrobp.2021.02.024

Zhang Y Y, Huo W L, Goldberg S I, Slater J M, Adams J A, Deng X W, Sun Y, Ma J, Fullerton B C, Paganetti H, Loeffler J S, Lu H M & Chan A W 2021 Brain-specific relative biological effectiveness of protons based on long-term outcome of patients with nasopharyngeal carcinoma. *International Journal of Radiation Oncology, Biology, Physics* **110**(4), 984–992.
  **URL:** http://dx.doi.org/10.1016/j.ijrobp.2021.02.018